\documentclass[longauth]{aa}

\usepackage{graphicx}
\usepackage{txfonts}
\usepackage{xspace}
\usepackage{color}
\usepackage{url}
\usepackage{hyperref}
\usepackage[flushleft]{threeparttable}
\newcommand {\bc}{\begin {center}}
\newcommand {\ec}{\end {center}}
\newcommand {\be}{\begin {equation}}
\newcommand {\ee}{\end {equation}}
\newcommand {\beq}{\begin {eqnarray}}
\newcommand {\eeq}{\end {eqnarray}}

\newcommand {\flux}{erg~s$^{-1}$~cm$^{-2}$}

\newcommand {\nustar}{\textit{NuSTAR}\xspace}
\newcommand {\chandra}{\textit{Chandra}\xspace}
\newcommand {\xrism}{\textit{XRISM}\xspace}
\newcommand {\ixpe}{\text{IXPE}\xspace}
\newcommand {\xmm}{\textit{XMM--Newton}\xspace}
\newcommand {\maxi}{\text{MAXI}\xspace}
\newcommand {\source}{{GX~13$+$1}\xspace}

\begin{document}

%\title{Detection of X-ray polarized emission from the galactic burster GX~13+1}
\title{Discovery of a strong rotation of the X-ray polarization angle in the galactic burster GX~13+1}

%preliminary author list
\author{
Anna Bobrikova \inst{\ref{in:UTU}}
\and Sofia V. Forsblom \inst{\ref{in:UTU}}
\and Alessandro Di Marco \inst{\ref{in:INAF-IAPS}}
\and Fabio La Monaca \inst{\ref{in:INAF-IAPS},\ref{in:UniRoma2},\ref{in:LaSapienza}} 
\and Juri Poutanen \inst{\ref{in:UTU}}
\and Mason Ng \inst{\ref{in:MIT}}
\and Swati Ravi \inst{\ref{in:MIT}}
\and Vladislav Loktev \inst{\ref{in:UTU}}
\and Jari J. E. Kajava \inst{\ref{in:Serco-ESA}}
\and Francesco Ursini \inst{\ref{in:UniRoma3}} 
\and Alexandra Veledina \inst{\ref{in:UTU},\ref{in:nordita}} 
\and Daniele Rogantini \inst{\ref{in:MIT}}
\and Tuomo~Salmi \inst{\ref{in:Amsterdam}}
\and Stefano Bianchi \inst{\ref{in:UniRoma3}} 
\and Fiamma Capitanio \inst{\ref{in:INAF-IAPS}}
\and Chris Done \inst{\ref{in:Durham},\ref{in:Tokyo}} 
%internal referee
\and Sergio     Fabiani \inst{\ref{in:INAF-IAPS}}
\and Andrea Gnarini \inst{\ref{in:UniRoma3}} 
\and \\ 
Jeremy Heyl \inst{\ref{in:UBC}} 
\and Philip Kaaret \inst{\ref{in:NASA-MSFC}}
%internal referee 
\and Giorgio Matt  \inst{\ref{in:UniRoma3}}  
\and Fabio Muleri \inst{\ref{in:INAF-IAPS}}  
\and Anagha P. Nitindala \inst{\ref{in:UTU}} 
\and John Rankin \inst{\ref{in:INAF-IAPS}} 
\and \\ 
Martin C. Weisskopf \inst{\ref{in:NASA-MSFC}} 
\and Iv\'an Agudo \inst{\ref{in:CSIC-IAA}}
\and Lucio A. Antonelli \inst{\ref{in:INAF-OAR},\ref{in:ASI-SSDC}} 
\and Matteo Bachetti \inst{\ref{in:INAF-OAC}} 
\and Luca~Baldini  \inst{\ref{in:INFN-PI},      \ref{in:UniPI}} 
\and \\ 
Wayne H. Baumgartner  \inst{\ref{in:NASA-MSFC}} 
\and Ronaldo Bellazzini  \inst{\ref{in:INFN-PI}}  
\and Stephen D. Bongiorno \inst{\ref{in:NASA-MSFC}} 
\and Raffaella Bonino  \inst{\ref{in:INFN-TO},\ref{in:UniTO}}
\and Alessandro Brez  \inst{\ref{in:INFN-PI}} 
\and Niccol\`{o} Bucciantini 
\inst{\ref{in:INAF-Arcetri},\ref{in:UniFI},\ref{in:INFN-FI}} 
\and Simone Castellano \inst{\ref{in:INFN-PI}}  
\and Elisabetta Cavazzuti \inst{\ref{in:ASI}} 
\and Chien-Ting Chen \inst{\ref{in:USRA-MSFC}}
\and Stefano Ciprini \inst{\ref{in:INFN-Roma2},\ref{in:ASI-SSDC}}
\and Enrico     Costa \inst{\ref{in:INAF-IAPS}} 
\and Alessandra De Rosa \inst{\ref{in:INAF-IAPS}} 
\and Ettore     Del Monte \inst{\ref{in:INAF-IAPS}} 
\and Laura      Di Gesu \inst{\ref{in:ASI}} 
\and Niccol\`{o} Di Lalla \inst{\ref{in:Stanford}}
\and \\
Immacolata      Donnarumma \inst{\ref{in:ASI}}
\and Victor Doroshenko \inst{\ref{in:Tub}}
\and Michal Dov\v{c}iak \inst{\ref{in:CAS-ASU}}
\and Steven     R. Ehlert \inst{\ref{in:NASA-MSFC}}  
\and Teruaki Enoto \inst{\ref{in:RIKEN}}
\and Yuri~Evangelista \inst{\ref{in:INAF-IAPS}}
\and Riccardo Ferrazzoli \inst{\ref{in:INAF-IAPS}} 
\and Javier     A. Garc\'{i}a \inst{\ref{in:GoddardXray}}
\and Shuichi Gunji \inst{\ref{in:Yamagata}} 
\and Kiyoshi Hayashida \inst{\ref{in:Osaka}} 
\and Wataru     Iwakiri \inst{\ref{in:Chiba}} 
\and Svetlana G. Jorstad \inst{\ref{in:BU},\ref{in:SPBU}}   
\and Vladimir Karas \inst{\ref{in:CAS-ASU}}
\and Fabian     Kislat \inst{\ref{in:UNH}} 
\and Takao      Kitaguchi  \inst{\ref{in:RIKEN}} 
\and Jeffery J. Kolodziejczak \inst{\ref{in:NASA-MSFC}} 
\and Henric~Krawczynski  \inst{\ref{in:WUStL}}
\and Luca Latronico  \inst{\ref{in:INFN-TO}} 
\and Ioannis Liodakis \inst{\ref{in:NASA-MSFC}}
\and Simone     Maldera \inst{\ref{in:INFN-TO}}  
\and Alberto Manfreda \inst{\ref{INFN-NA}}
\and Fr\'{e}d\'{e}ric~Marin \inst{\ref{in:Strasbourg}} 
\and Andrea     Marinucci \inst{\ref{in:ASI}} 
\and Alan P. Marscher \inst{\ref{in:BU}} 
\and  Herman L. Marshall \inst{\ref{in:MIT}}
\and Francesco  Massaro \inst{\ref{in:INFN-TO},\ref{in:UniTO}} 
\and Ikuyuki~Mitsuishi \inst{\ref{in:Nagoya}} 
\and Tsunefumi  Mizuno \inst{\ref{in:Hiroshima}} 
\and Michela Negro \inst{\ref{in:LSU}} 
\and Chi-Yung Ng \inst{\ref{in:HKU}}
\and Stephen L. O'Dell \inst{\ref{in:NASA-MSFC}}  
\and Nicola     Omodei \inst{\ref{in:Stanford}}
\and Chiara     Oppedisano \inst{\ref{in:INFN-TO}}  
\and Alessandro Papitto \inst{\ref{in:INAF-OAR}}
\and George     G. Pavlov \inst{\ref{in:PSU}}
\and Abel L. Peirson \inst{\ref{in:Stanford}}
\and Matteo     Perri \inst{\ref{in:ASI-SSDC},\ref{in:INAF-OAR}}
\and Melissa~Pesce-Rollins \inst{\ref{in:INFN-PI}} 
\and Pierre-Olivier     Petrucci \inst{\ref{in:Grenoble}} 
\and Maura Pilia \inst{\ref{in:INAF-OAC}} 
\and Andrea     Possenti \inst{\ref{in:INAF-OAC}} 
\and Simonetta  Puccetti \inst{\ref{in:ASI-SSDC}}
\and Brian~D.~Ramsey \inst{\ref{in:NASA-MSFC}}  
\and Ajay Ratheesh \inst{\ref{in:INAF-IAPS}} 
\and Oliver     J. Roberts \inst{\ref{in:USRA-MSFC}}
\and Roger W. Romani \inst{\ref{in:Stanford}}
\and Carmelo Sgr\`o \inst{\ref{in:INFN-PI}}  
\and Patrick Slane \inst{\ref{in:CfA}}  
\and Paolo~Soffitta \inst{\ref{in:INAF-IAPS}} 
\and Gloria     Spandre \inst{\ref{in:INFN-PI}} 
\and Douglas A. Swartz \inst{\ref{in:USRA-MSFC}}
\and Toru Tamagawa \inst{\ref{in:RIKEN}}
\and Fabrizio Tavecchio \inst{\ref{in:INAF-OAB}}
\and Roberto Taverna \inst{\ref{in:UniPD}} 
\and Yuzuru     Tawara \inst{\ref{in:Nagoya}}
\and Allyn F. Tennant \inst{\ref{in:NASA-MSFC}}  
\and Nicholas E. Thomas \inst{\ref{in:NASA-MSFC}}  
\and Francesco  Tombesi  \inst{\ref{in:UniRoma2},\ref{in:INFN-Roma2},\ref{in:UMd}}
\and Alessio Trois \inst{\ref{in:INAF-OAC}}
\and Sergey~S.~Tsygankov \inst{\ref{in:UTU}}
\and Roberto Turolla \inst{\ref{in:UniPD},\ref{in:MSSL}}
\and Jacco Vink \inst{\ref{in:Amsterdam}}
\and Kinwah     Wu \inst{\ref{in:MSSL}}
\and Fei Xie \inst{\ref{in:GSU},\ref{in:INAF-IAPS}}
\and Silvia Zane  \inst{\ref{in:MSSL}}
          }
          
\institute{
Department of Physics and Astronomy, FI-20014 University of Turku,  Finland \label{in:UTU} \\ 
\email{anna.a.bobrikova@utu.fi} 
\and 
INAF Istituto di Astrofisica e Planetologia Spaziali, Via del Fosso del Cavaliere 100, 00133 Roma, Italy \label{in:INAF-IAPS}
\and
Dipartimento di Fisica, Universit\`{a} degli Studi di Roma ``Tor Vergata'', Via della Ricerca Scientifica 1, 00133 Roma, Italy \label{in:UniRoma2} 
\and 
Dipartimento di Fisica, Universit\`{a} degli Studi di Roma ``La Sapienza'', Piazzale Aldo Moro 5, 00185 Roma, Italy \label{in:LaSapienza}
\and
MIT Kavli Institute for Astrophysics and Space Research, Massachusetts Institute of Technology, 77 Massachusetts Avenue, Cambridge, MA 02139, USA \label{in:MIT}
\and 
Serco for the European Space Agency (ESA), European Space Astronomy Centre, Camino Bajo del Castillo s/n, E-28692 Villanueva de la Ca\~{n}ada, Madrid, Spain \label{in:Serco-ESA} 
\and 
Dipartimento di Matematica e Fisica, Universit\`a degli Studi Roma Tre, via della Vasca Navale 84, 00146 Roma, Italy  \label{in:UniRoma3}
\and 
Nordita, KTH Royal Institute of Technology and Stockholm University, Hannes Alfv{\'e}ns v{\"a}g 12, SE-106 91, Sweden
\label{in:nordita}
\and
Anton Pannekoek Institute for Astronomy \& GRAPPA, University of Amsterdam, Science Park 904, 1098 XH Amsterdam, The Netherlands  \label{in:Amsterdam}
\and 
Centre for Extragalactic Astronomy, Department of Physics, Durham University, South Road, Durham DH1 3LE, UK \label{in:Durham} 
\and 
Kavli Institute for Physics and Mathematics of the Universe (WPI), University of Tokyo, Kashiwa, Chiba 277-8583, Japan \label{in:Tokyo} 
\and 
University of British Columbia, Vancouver, BC V6T 1Z4, Canada \label{in:UBC}
\and 
NASA Marshall Space Flight Center, Huntsville, AL 35812, USA \label{in:NASA-MSFC}
\and 
Instituto de Astrof\'{i}sica de Andaluc\'{i}a -- CSIC, Glorieta de la Astronom\'{i}a s/n, 18008 Granada, Spain \label{in:CSIC-IAA}
\and 
INAF Osservatorio Astronomico di Roma, Via Frascati 33, 00040 Monte Porzio Catone (RM), Italy \label{in:INAF-OAR} 
\and 
Space Science Data Center, Agenzia Spaziale Italiana, Via del Politecnico snc, 00133 Roma, Italy \label{in:ASI-SSDC}
 \and
INAF Osservatorio Astronomico di Cagliari, Via della Scienza 5, 09047 Selargius (CA), Italy  \label{in:INAF-OAC}
\and 
Istituto Nazionale di Fisica Nucleare, Sezione di Pisa, Largo B. Pontecorvo 3, 56127 Pisa, Italy \label{in:INFN-PI}
\and  
Dipartimento di Fisica, Universit\`{a} di Pisa, Largo B. Pontecorvo 3, 56127 Pisa, Italy \label{in:UniPI} 
\and  
Istituto Nazionale di Fisica Nucleare, Sezione di Torino, Via Pietro Giuria 1, 10125 Torino, Italy  \label{in:INFN-TO}      
\and  
Dipartimento di Fisica, Universit\`{a} degli Studi di Torino, Via Pietro Giuria 1, 10125 Torino, Italy \label{in:UniTO} 
\and   
INAF Osservatorio Astrofisico di Arcetri, Largo Enrico Fermi 5, 50125 Firenze, Italy 
\label{in:INAF-Arcetri} 
\and  
Dipartimento di Fisica e Astronomia, Universit\`{a} degli Studi di Firenze, Via Sansone 1, 50019 Sesto Fiorentino (FI), Italy \label{in:UniFI} 
\and   
Istituto Nazionale di Fisica Nucleare, Sezione di Firenze, Via Sansone 1, 50019 Sesto Fiorentino (FI), Italy \label{in:INFN-FI}
\and 
Agenzia Spaziale Italiana, Via del Politecnico snc, 00133 Roma, Italy \label{in:ASI}
\and 
Science and Technology Institute, Universities Space Research Association, Huntsville, AL 35805, USA \label{in:USRA-MSFC}
\and 
Istituto Nazionale di Fisica Nucleare, Sezione di Roma ``Tor Vergata'', Via della Ricerca Scientifica 1, 00133 Roma, Italy 
\label{in:INFN-Roma2}
\and 
Department of Physics and Kavli Institute for Particle Astrophysics and Cosmology, Stanford University, Stanford, California 94305, USA  \label{in:Stanford}
\and
Institut f\"ur Astronomie und Astrophysik, Universit\"at T\"ubingen, Sand 1, D-72076 T\"ubingen, Germany \label{in:Tub}
\and 
Astronomical Institute of the Czech Academy of Sciences, Bo\v{c}n\'{i} II 1401/1, 14100 Praha 4, Czech Republic \label{in:CAS-ASU}
\and 
RIKEN Cluster for Pioneering Research, 2-1 Hirosawa, Wako, Saitama 351-0198, Japan \label{in:RIKEN}
\and 
X-ray Astrophysics Laboratory, NASA Goddard Space Flight Center, Greenbelt, MD 20771, USA \label{in:GoddardXray}
\and 
Yamagata University,1-4-12 Kojirakawa-machi, Yamagata-shi 990-8560, Japan \label{in:Yamagata}
\and 
Osaka University, 1-1 Yamadaoka, Suita, Osaka 565-0871, Japan \label{in:Osaka}
\and 
International Center for Hadron Astrophysics, Chiba University, Chiba 263-8522, Japan \label{in:Chiba}
\and
Institute for Astrophysical Research, Boston University, 725 Commonwealth Avenue, Boston, MA 02215, USA \label{in:BU} 
\and 
Department of Astrophysics, St. Petersburg State University, Universitetsky pr. 28, Petrodvoretz, 198504 St. Petersburg, Russia \label{in:SPBU} 
\and 
Department of Physics and Astronomy and Space Science Center, University of New Hampshire, Durham, NH 03824, USA \label{in:UNH} 
\and 
Physics Department and McDonnell Center for the Space Sciences, Washington University in St. Louis, St. Louis, MO 63130, USA \label{in:WUStL}
\and 
Istituto Nazionale di Fisica Nucleare, Sezione di Napoli, Strada Comunale Cinthia, 80126 Napoli, Italy \label{INFN-NA}
\and 
Universit\'{e} de Strasbourg, CNRS, Observatoire Astronomique de Strasbourg, UMR 7550, 67000 Strasbourg, France \label{in:Strasbourg}
\and 
Graduate School of Science, Division of Particle and Astrophysical Science, Nagoya University, Furo-cho, Chikusa-ku, Nagoya, Aichi 464-8602, Japan \label{in:Nagoya}
\and 
Hiroshima Astrophysical Science Center, Hiroshima University, 1-3-1 Kagamiyama, Higashi-Hiroshima, Hiroshima 739-8526, Japan\label{in:Hiroshima}
\and 
Department of Physics and Astronomy, Louisiana State University, Baton Rouge, LA 70803, USA \label{in:LSU}
\and 
Department of Physics, University of Hong Kong, Pokfulam, Hong Kong \label{in:HKU}
\and 
Department of Astronomy and Astrophysics, Pennsylvania State University, University Park, PA 16801, USA \label{in:PSU}
\and 
Universit\'{e} Grenoble Alpes, CNRS, IPAG, 38000 Grenoble, France \label{in:Grenoble}
\and 
Center for Astrophysics, Harvard \& Smithsonian, 60 Garden St, Cambridge, MA 02138, USA \label{in:CfA} 
\and 
INAF Osservatorio Astronomico di Brera, via E. Bianchi 46, 23807 Merate (LC), Italy \label{in:INAF-OAB}
\and 
Dipartimento di Fisica e Astronomia, Universit\`{a} degli Studi di Padova, Via Marzolo 8, 35131 Padova, Italy \label{in:UniPD}
\and
Department of Astronomy, University of Maryland, College Park, Maryland 20742, USA \label{in:UMd}
\and 
Mullard Space Science Laboratory, University College London, Holmbury St Mary, Dorking, Surrey RH5 6NT, UK \label{in:MSSL}
\and 
Guangxi Key Laboratory for Relativistic Astrophysics, School of Physical Science and Technology, Guangxi University, Nanning 530004, China \label{in:GSU}
}
          
\titlerunning{First detection of polarized emission from \source}
\authorrunning{Bobrikova et al.}

\date{2024}

\abstract
{Weakly magnetized neutron stars in X-ray binaries show a complex phenomenology with several spectral components that can be associated with the accretion disk, the boundary, and/or a spreading layer, a corona, and a wind. 
Spectroscopic information alone, however, is not enough to distinguish these components. 
The analysis of the timing data revealed that most of the variability, and in particular, kilohertz quasi-period oscillations, are associated with the high-energy component that corresponds to the boundary and/or spreading layer. 
Additional information about the nature of the spectral components, and in particular, about the geometry of the emission region, can be provided by X-ray polarimetry. 
One of the objects of the class, a bright, persistent, and rather peculiar galactic Type I X-ray burster \source, was observed with the \textit{Imaging X-ray Polarimetry Explorer} (\ixpe) and the \xmm. 
Using the \xmm data, we obtained the best-fit values for the continuum spectral parameters and detected strong absorption lines associated with the accretion disk wind. 
\ixpe data showed the source to be significantly polarized in the 2--8 keV energy band, with an overall polarization degree (PD) of $1.4\%\pm0.3\%$ at a polarization angle (PA) of $-2\degr\pm6\degr$ (errors at the 68\% confidence level). 
During the two-day long observation, we detected rotation of the PA by about 70\degr\ with the corresponding changes in the PD from 2\% to nondetectable and then up to 5\%. 
These variations in polarization properties are not accompanied by visible spectral state changes of the source.
The energy-resolved polarimetric analysis showed a significant change in polarization, from being strongly dependent on energy at the beginning of the observation to being almost constant with energy in the later parts of the observation. 
As a possible interpretation, we suggest a constant polarization component, strong wind scattering, or a different polarization of the two main spectral components with an individually peculiar behavior. 
The rotation of the PA suggests a %30\degr--40\degr\ 
misalignment of the neutron star spin from the orbital axis. 
}

\keywords{accretion, accretion disks -- polarization -- stars: neutron -- X-rays: binaries}

\maketitle

%%%%%%%%%%%%%%%%%%%%%%%%%%%%%%%%%%%%%%%%%%%%%%%%%%%%%%%%%%%%%%%%%%%%%%%%%%%%%%
\section{Introduction} 
\label{sec:intro}

Weakly magnetized neutron stars (WMNSs) reside in low-mass X-ray binary (LMXB) systems and accrete material from companion stars via an accretion disk. 
With their luminosity ranging in the $10^{36}-10^{38}$ erg\,s$^{-1}$, they are among the brightest X-ray sources in the sky.
WMNSs can be classified as Z- or atoll-sources, based on the shape of their tracks in the X-ray hardness-intensity (HID) diagram or in the color-color diagram (CCD), their timing proprieties in the 1--10 keV band, their radio emission, and their luminosity and mass-accretion rate. WMNSs are also known for their fast X-ray variability: The quasi-periodic oscillations in the ranges seconds and milliseconds that were observed from many sources of the class \citep{vdKlis1989,vdKlis2000,2021ASSL..461..263M} are associated with a harder spectral component \citep{2003A&A...410..217G,RG2006,Revnivtsev2013} close to the neutron star (NS) surface.

In the absence of a strong magnetic field surrounding the NS in these systems, the matter from the accretion disk falls onto the equator of the star via the boundary layer \citep[BL, see e.g.][]{1988AdSpR...8b.135S} and further spreads throughout the NS surface, where it forms the so-called spreading layer \citep[SL, see e.g.][]{1985MNRAS.217..291L, IS1999}. 
The geometry and physics of the processes that occur within these layers both affect the emission of WMNSs and their polarization \citep[see e.g. ][]{1985MNRAS.217..291L, Gnarini22}.
The emission of WMNSs in the X-ray band has two main components: a soft thermal emission, which can be blackbody-like emission from the NS surface or a multicolor emission from the disk, and a harder component associated with the Comptonization in the BL or SL. 
In addition to these two main components, signs of reflection of the SL emission from the accretion disk (usually a broad emission line around 6.4 keV) and a set of absorption lines are sometimes visible, which indicates that the radiation interacts in the wind above the disk. 
A contribution of the hot corona is also possible, as is a contribution of the hot flow in the hard state.

Polarimetry is a useful tool for understanding the geometry of the source and the emission mechanisms by studying the two additional variables it provides: the polarization degree (PD), and polarization angle (PA). 
The Imaging X-ray Polarimetry Explorer (\ixpe; \citealt{Weisskopf2022}), launched in December 2021, is a pathfinder that gives us the unique opportunity to measure the X-ray polarimetric properties of WMNSs with high precision. In the last two years, nine objects of the class were observed \citep[see][]{Capitanio23, Farinelli23, Cocchi23, Ursini2023, DiMarco23, Rankin2024, LaMonaca2024, Saade24,Fabiani24}. 
These observations shed new light on the emission mechanisms of these sources, as well as on the changes in PD and PA while the source changes its spectral state. 
We aim to add another important piece of evidence to the global picture. % by focusing on the radiation scattering in the wind above the accretion disk.

\source is a peculiar source with a unique position between atolls and Z-sources. It is located at a distance of $7\pm1$ kpc in a binary system and has a late-evolved K5 III giant as a companion \citep{1999MNRAS.306..417B}. 
The source shows several features that have traditionally been associated with Z-sources: It shines almost as brightly as Z-sources \citep[up to $0.5 L_{\rm Edd}$,][]{Dai14}, it has persistent radio emission \citep{Grindlay86}, and it shows 57--69 Hz quasi-periodic oscillations \citep{Homan98}. 
The path it draws on the CCD, however, is closer to the path normally observed from the atolls \citep{2003A&A...406..221S}. 
However, even the CCD of the source has recently been interpreted as if \source were a Z-source \citep{2023MNRAS.522.3367S, 2018ApJ...861...26A}. 
Another unique property of \source is the complex spectrum \citep[e.g.,][]{DT12}. 
In addition to the common softer accretion disk and harder Comptonized components, a sign of reflection of the Comptonized emission from the disk is present, as well as seven absorption features caused by the interaction of the radiation in the wind above the disk. 
\citet{2018ApJ...861...26A} showed that these absorption features could not be produced by a single absorbing region, which indicates a complicated structure of the wind.
\source also exhibits periodic variations in the emission with a period of 24.5~d \citep{2010ApJ...719..979C}. 
These modulations are associated with the orbital movement of the source, making \source the NS--LMXB with the longest orbital period. 
Short dips of $\approx$1~ks in the X-ray spectrum are associated with this orbital movement \citep{2010AIPC.1248..153D, 2014A&A...561A..99I}, but they are occasional events and are not present in every orbit. 
These dips are traditionally interpreted as an indication of the high source inclination ($60\degr-75\degr$). They come from the interaction of the emission of the NS with the mostly neutral accretion bulge located in the outer parts of the disk \citep{1982ApJ...257..318W,Dai14}.

The wind above the accretion disk of \source has been studied with several observatories. 
For instance, using \chandra data, \citet{2020MNRAS.497.4970T} gave constraints on the radial and azimuthal velocity of the wind, and \citet{2018ApJ...861...26A} constrained the wind-launching radius, showing that the radiation pressure likely launches the wind. 
\citet{2019A&A...625A...8M} developed a nonrelativistic framework capable of fitting and interpreting the asymmetric K$\alpha$ lines in \xmm data and reported that this nonrelativistic approach performs as well as the known relativistic framework. This adds an alternative explanation of the red-skewed Fe line in the spectrum of \source. 
\citet{2023MNRAS.522.3367S} used \nustar data to estimate the distance between the source and the ionized absorbing material of $(4-40)\times10^5$~km, constrained the magnetic field at $B\lesssim1.8\times10^8$~G, and pointed out the overabundance of Fe and Ni in the source. 
\source is also in the list of targets for the recently launched \textit{X-ray Imaging and Spectroscopy Mission} \citep[\xrism;][]{XRISM}. 
The aim of our project is to obtain constraints on the source geometry and radiation mechanisms using the polarimetric capabilities of \ixpe.  
%We contribute to the ongoing study of this source with a polarimetric view and with a result of complimentary spectroscopic observation. 

The remainder of the paper is structured as follows. We introduce the observations performed with \ixpe and the \xmm observatories in Sect.~\ref{sec:obs}. 
In Sect.~\ref{sec:analysis} we present the data analysis. 
We provide an interpretation of the results in Sect.~\ref{sec:discussion} and summarize in Sect.~\ref{sec:summary}.

\section{Observations} 
\label{sec:obs}

\begin{table*}
\centering
\caption{Observations of GX 13+1 presented in the paper.}
\label{table:obsdates}
\begin{tabular}{ccccc}
            \hline\hline
%\noalign{\smallskip}
 Observatory & Dates & ObsID & Instrument & Duration (s) \\
%\noalign{\smallskip}
\hline
%\noalign{\smallskip}
            \ixpe & 2023 Oct 17--19 & 02006801 & DU1 & 98204     \\
              & &  & DU2 & 98299 \\
              &  &  & DU3 & 98299 \\
            \xmm & 2023 Oct 16 & 0932390601 & EPIC/PN & 11336 \\
%              &  &  RGS1 & 13395 \\
%               & & RGS2 & 13390 \\
%               & & OM & 4100             \\
%\noalign{\smallskip}
            \hline
         \end{tabular}
   \end{table*}

\subsection{IXPE}

The \ixpe was launched on 2021 December 9 as a result of joint efforts of NASA and the Italian Space Agency. 
It has three identical grazing-incidence telescopes on board, each of which comprises an X-ray mirror assembly and a polarization-sensitive detector unit (DU) equipped with a gas-pixel detector \citep{2021AJ....162..208S,2021APh...13302628B}. 
\ixpe is capable of performing imaging polarimetry over the 2--8 keV energy band with a time resolution of about 10~$\mu$s. 
A detailed description of the observatory and its performance is given in \citet{Weisskopf2022}.

\begin{figure}
\centering
\includegraphics[width = 0.85\linewidth]{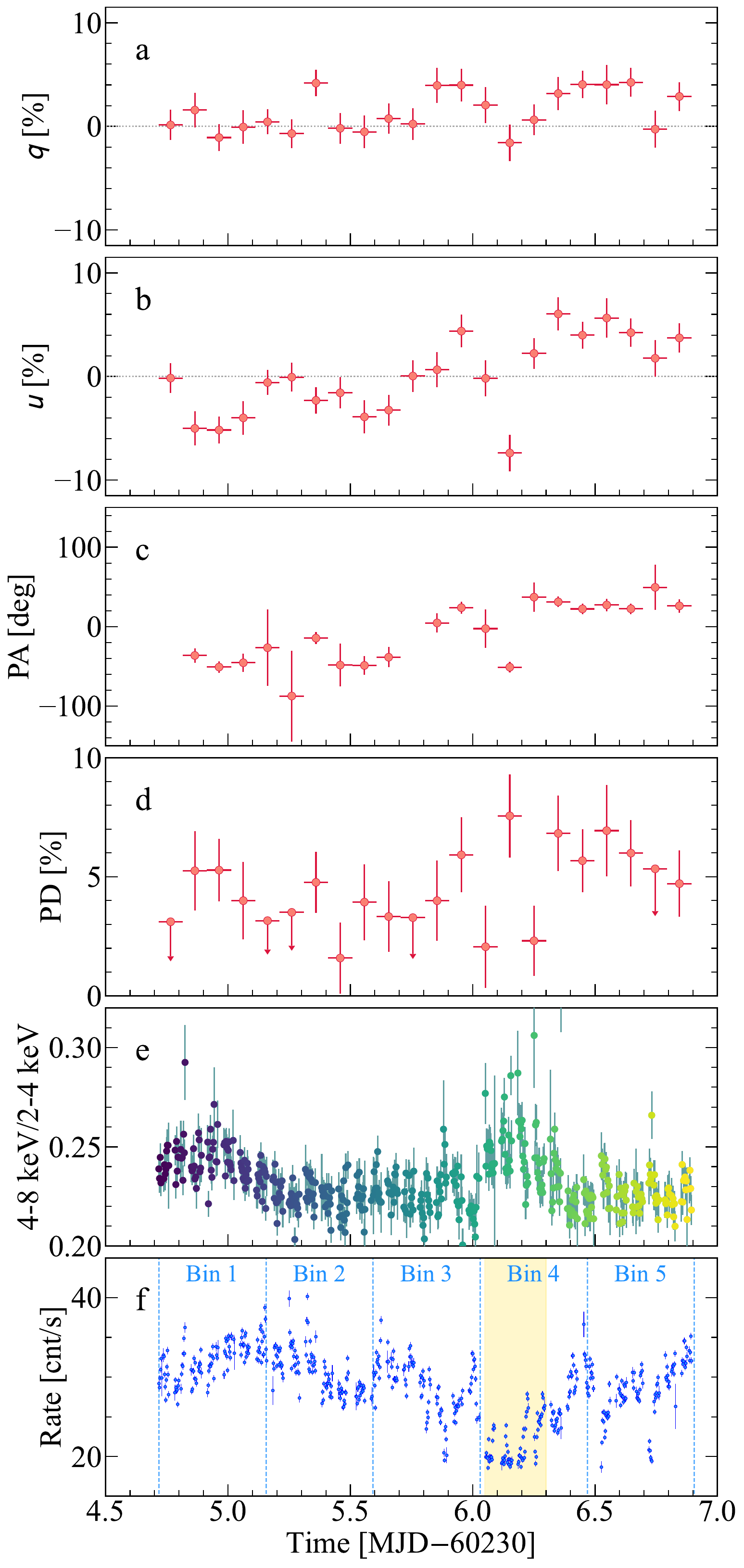}
\caption{Normalized Stokes $q$ (a) and $u$ (b) parameters, PA (c), PD (d), hardness ratio (e) with the same colors as in Fig.~\ref{fig:HID}, and the count rate (f) as functions of time as observed by \ixpe (all three detectors combined in the energy range 2--8 keV). 
The light curve is binned in $\approx$200~s. The dashed vertical blue lines separate the observation into five equal time bins of 10.5~h (Bins~1--5). 
The region highlighted in yellow corresponds to the dip in the light curve (see Sect.~\ref{sec:PCUBE}). 
Uncertainties are reported at the $68\%$ CL. }
\label{fig:lightcurve}%
\end{figure}

\ixpe observed \source on 2023 October 17--19 with a total effective exposure of $\simeq$100~ks for each of the three telescopes (see Table~\ref{table:obsdates}). 
For the data analysis, we used the \textsc{ixpeobssim} software, version 30.6.3 \citep{Baldini2022}, and the \textsc{heasoft} package, version 6.31.1 \citep{heasarc}. 
We used the latest version of the \ixpe response matrices, released at the \textsc{Heasarc} \texttt{CALDB} on 2023 June 16. 
We extracted the $I$, $Q$, and $U$ spectra from a region with a radius of 80\arcsec. 
Following \citet{DiMarco2023}, we did not perform background subtraction or rejection, as \source is sufficiently bright for the background to be negligible. 
The polarimetric analysis was performed using the formalism of \citet{2015APh....68...45K} implemented in the \textsc{ixpeobssim} package \citep{Baldini2022} under the \texttt{pcube} algorithm in the \texttt{xpbin} tool. 
We used \textsc{ixpeobssim} to investigate the time-dependence of the normalized Stokes parameters $q=Q/I$ and $u=U/I$, as well as the PD$=\sqrt{q^2+u^2}$  and PA$=\frac{1}{2}\arctan(u/q)$ (see Fig.~\ref{fig:lightcurve}a-d). We applied the unweighted analysis implemented in the \textsc{ixpeobssim}.
For the spectral and spectropolarimetric analysis, we used \textsc{xspec}, version 12.13.0c \citep{Arnaud96}, and applied a weighted analysis as introduced in \citet{DiMarco2022}. 
For the spectral analysis, the data were binned to have at least 30~counts in each energy bin. 
For the spectropolarimetric analysis, we applied a constant energy binning of 80~eV for the three Stokes parameters.

\begin{figure}
\centering
\includegraphics[width=0.9\linewidth]{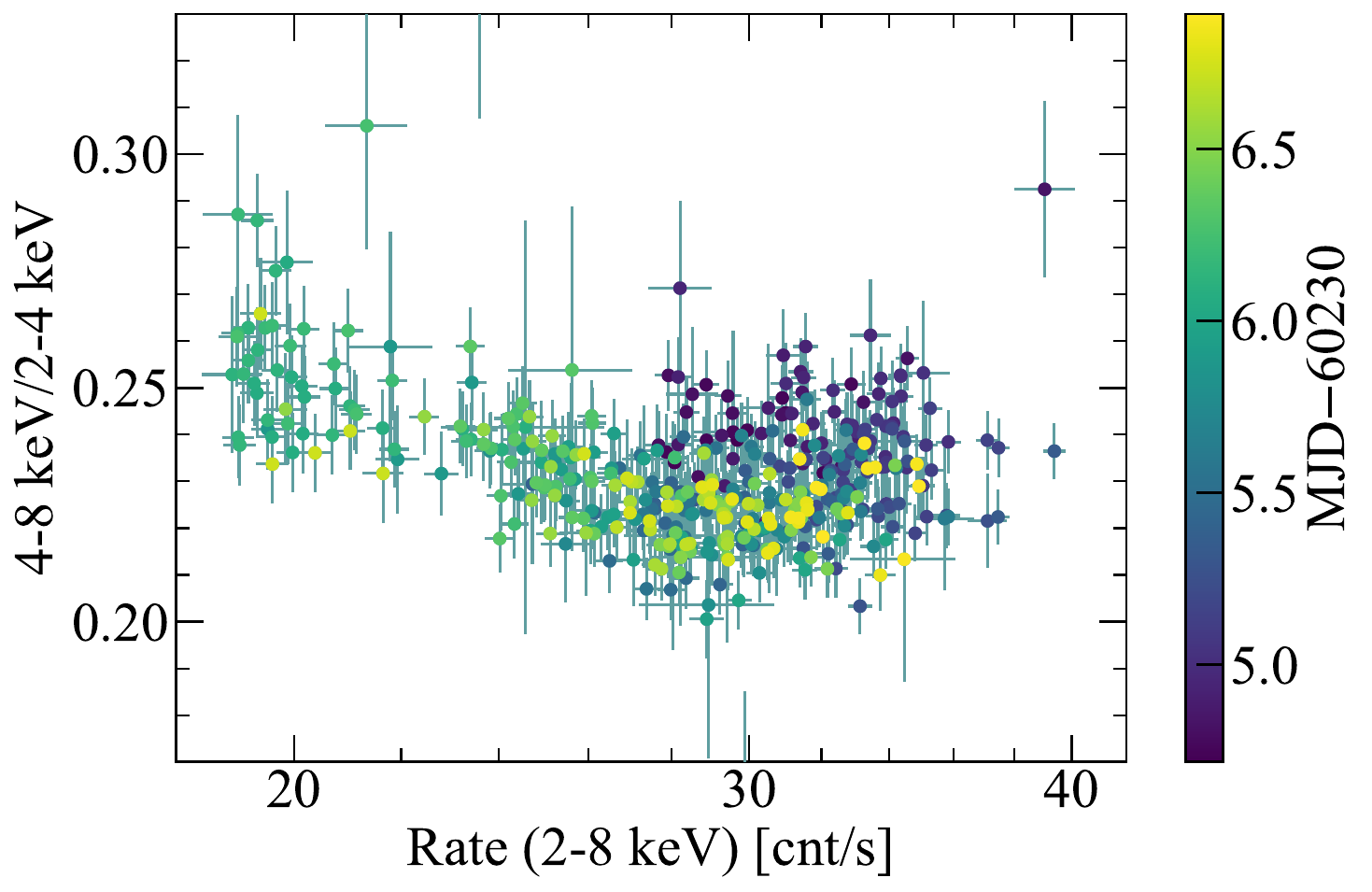}
\caption{Hardness-intensity diagram derived from the \ixpe data in time bins of 200~s. 
The colors correspond to the evolution of the data with time, from dark blue at the beginning of the observation to yellow at the end of the observation. }
\label{fig:HID}
\end{figure}

\subsection{\xmm}

\xmm observed \source on 2023 October 16 (see Table~\ref{table:obsdates}). This observation could not be scheduled simultaneously with \ixpe because the visibility window for \source with \xmm closed on October 16. 
As \source is very bright, EPIC/PN was operated in timing mode. The thick filter was used to reduce the count rate, which was about 690\,cnt\,s$^{-1}$ on average during the observation. 
The EPIC/MOS instruments were not used, and their telemetry was allocated to EPIC/PN. 
Despite this effort, PN nevertheless suffered from telemetry gaps as the science buffer was full throughout the observation. 
For this reason, the data are not suitable for a timing analysis, but are still useful for measuring the spectral shape and the known wind lines in the 6--9~keV range \citep{DT12}.

The data were reduced using \textsc{xmm--sas} v.21 with the latest calibration files as of October 2023.
The clean event files were generated from the ODF files with the \texttt{epchain} tool. 
They were filtered further using the \#XMMEA\_EP and PATTERN$==$0 flags in the 0.3--12 keV range. 
No soft proton flares were seen in the light curves above 10 keV, giving us a total effective exposure of $\simeq$10~ks.
To extract the source spectrum, we selected events from a 9-pixel wide box around the RAWX column with the highest count rate.
We also extracted a background spectrum from RAWX columns 3--5, but as expected for such a bright target, these columns were still dominated by the source itself, and thus, we did not subtract the background at all.
The ancillary response file and the redistribution matrix were generated using the tools \texttt{arfgen} and \texttt{rmfgen}, respectively.
Finally, we grouped the data to have at least 20 counts per channel and added  $1\%$ systematic errors in quadrature to each channel using the \texttt{grppha} tool.

\section{Data analysis}
\label{sec:analysis}

\subsection{Model-independent polarimetric analysis} \label{sec:PCUBE}

The light curve of \source is shown in Fig.~\ref{fig:lightcurve}f. 
The source showed some variations, with the count rate dropping by almost a factor of two in the wide dip, but some more rapid dips are noticeable.  
The hardness ratio (Fig.~\ref{fig:lightcurve}e) changes with time by only a small margin; the largest increase slightly after MJD 60236 coincides with the wide dip in the light curve and is $\approx10\%$.  

The significant variability of the Stokes parameters is immediately obvious. 
This variability is reflected in the changes in PD and PA in Fig.~\ref{fig:lightcurve}c,d. 
While the PA slowly increases throughout the observation, the PD changes significantly and without an obvious pattern. 

\begin{figure}
\centering
\includegraphics[width = 0.9\linewidth]{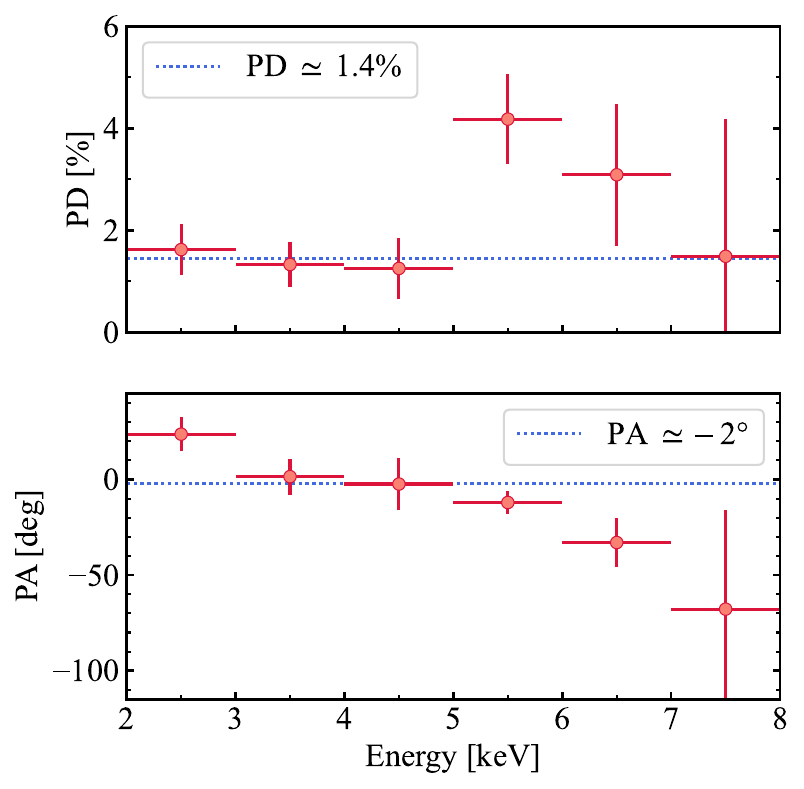}
\caption{Energy-resolved PD and PA averaged over all the \ixpe observation, obtained with the \texttt{pcube} algorithm. 
The data are divided into six 1-keV wide energy bins. 
Uncertainties are reported at the $68\%$ CL. }
\label{fig:pdpa_ene_ave}
\end{figure}

The HID presented in Fig.~\ref{fig:HID} illustrates a lack of state variability within the more than two days of \ixpe observation. 
From the data publicly available in the \maxi \citep{MAXI} archive\footnote{ \url{https://maxi.riken.jp/v7l3h/J1814-171/index.html}} of the most recent observations, we calculated the CCD from the ratio of fluxes $F_{10-20\,\rm keV}/F_{4-10\,\rm keV}$  against the ratio $F_{4-10\,\rm  keV}/F_{2-4\,\rm keV}$. We then estimated the position of the points corresponding to the observation days on the CCD and concluded that the source during the \ixpe observation was in the so-called lower left banana (LLB) state, a soft state common for the atoll sources. 
In the two days of \ixpe observation, \source moved in the lower left part of the CCD, but the uncertainties are too high to conclude about the state variability. 
Polarimetric analysis of the averaged data in the 2--8 keV energy band resulted in low PD values of $1.4\%\pm0.3\%$ and a PA of $-2\degr\pm6\degr$ (from here onward, the errors are given at the $68\%$ confidence level, CL) with the significance of the detection exceeding $99.99\%$  (see Fig.~\ref{fig:pdpa_ene_ave} and Table~\ref{tab:PCUBE}).

\begin{table} % [h!]
\centering
\caption{Results of the \texttt{pcube} analysis performed with various approaches to separate the data.} 
%\scriptsize
\footnotesize
\begin{tabular}{lcc }
\hline
\hline
Part of the observation & PD (\%) & PA (deg)  \\
\hline 
overall & $1.4\pm0.3$ & $-2\pm6$\\
\hline
pre-dip & $1.8\pm0.4$ & $-29\pm6$\\
dip & $2.2\pm1.1$ & $-49\pm14$\\
post-dip & $5.3\pm0.6$ & $26\pm3$\\
dip-core & $7.5\pm1.7$ & $-51\pm7$\\
\hline
Bin~1 & $3.2\pm0.7$ & $-44\pm6$\\
Bin~2 & $1.5\pm0.7$ & $-21\pm13$\\
Bin~3 & $1.6\pm0.7$ & $4\pm13$\\
Bin~4 & $1.6\pm0.7$ & $12\pm13$\\
Bin~5 & $5.0\pm0.7$ & $26\pm4$\\
\hline
Dimmer state, <25\,cnt\,s$^{-1}$ & $1.3\pm0.9$ & $-36\pm18$\\
Brighter state, >25\,cnt\,s$^{-1}$ & $1.6\pm0.3$ & $2\pm6$\\
\hline
\end{tabular}
\tablefoot{The errors are at 68\% CL. The results are presented for the overall polarization during the observation, separation by the light curve into pre-dip, dip, and post-dip parts with a special mention of the dip-core, one bin at $\approx6.15$~d in Fig.~\ref{fig:lightcurve}, separation into five equal time bins, and separation by the count rate into dimmer and brighter states. }  
\label{tab:PCUBE}
\end{table}  

\begin{figure}
\centering
\includegraphics[width = 1.\linewidth]{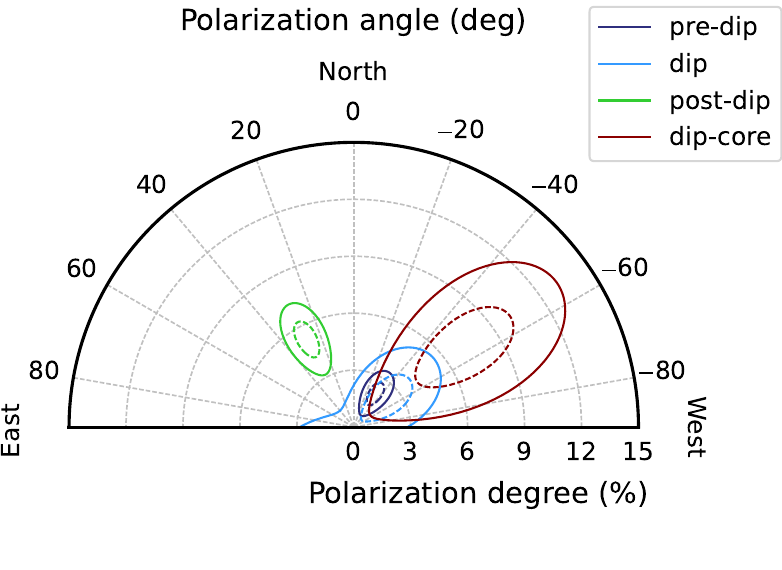}
\caption{Polar plot of the PD and PA in the 2--8\,keV energy band for the three parts of the observation (pre-dip, dip, and post-dip) and the dip core. 
Contours are at the $68\%$ CL (dashed lines) and $99\%$ CL (solid lines). }
\label{fig:contour_three}
\end{figure}

As the HID does not show any significant change of state, we decided to perform a more detailed study of different parts of the observation based on the light curve. 
We separated the dip state (highlighted in yellow in Fig.~\ref{fig:lightcurve}f) from the nondip states and divided the observation into three unequal parts: the pre-dip (before $t=6.05$~d), the dip ($6.05<t<6.30$~d), and the post-dip ($t> 6.30$~d), with the times $t$ given relative to MJD 60230. 
The results are gathered in Fig.~\ref{fig:contour_three}. 
The PD in the dip state is $2.2\%\pm1.1\%$, and the PA is $-49\degr \pm14\degr$ at $68\%$ CL. 
Due to the lack of statistics, we chose not to perform an energy-resolved analysis of this part of the observation. 
 
\begin{figure*}
\centering
\includegraphics[width = 0.4\linewidth]{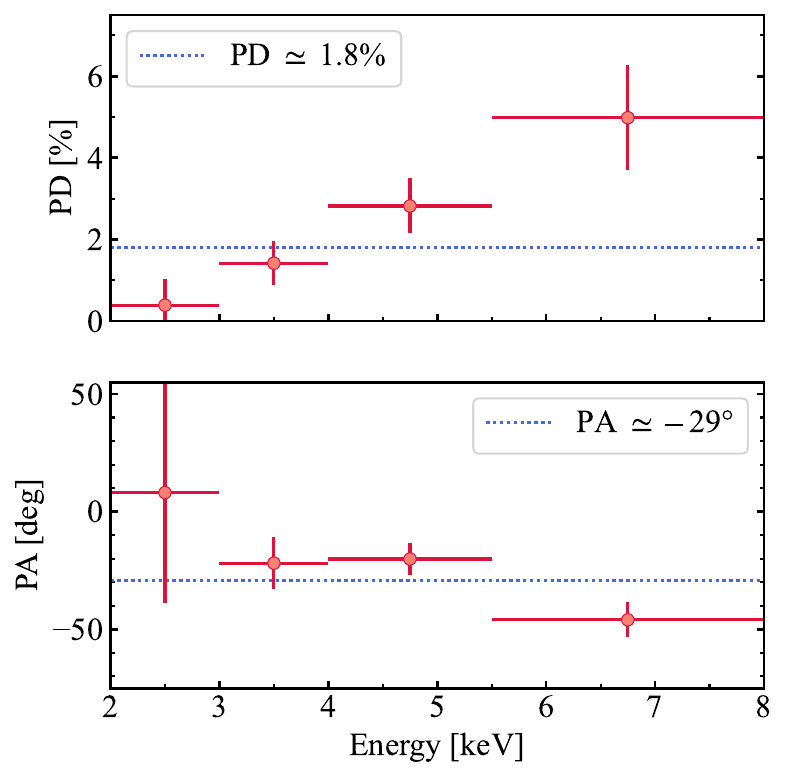}
\includegraphics[width = 0.4\linewidth]{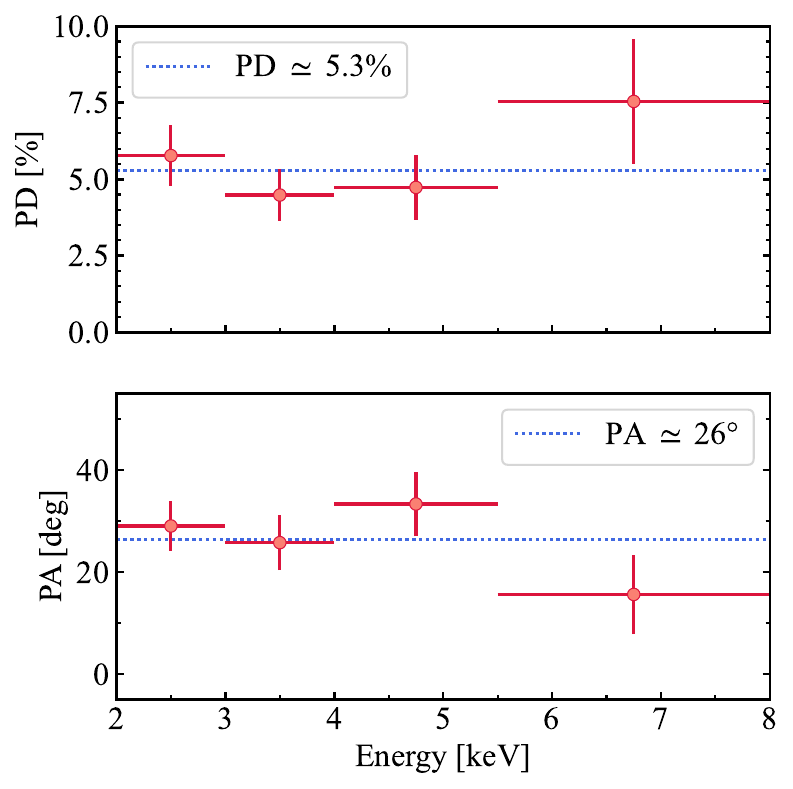}
\caption{Energy dependence of the PD and PA obtained by the \texttt{pcube} algorithm. 
Left: Pre-dip part of the observation. Right: Post-dip part of the observation. 
%Separation of the observation is performed based on the light curve shown in Fig.~\ref{fig:lightcurve}f. 
The data are divided into four energy bins: 2--3, 3--4, 4--5.5, and 5.5--8 keV. 
Uncertainties are reported at $68\%$ CL.}
\label{fig:pdpa_ene_split}
\end{figure*}

\begin{figure*}
\centering
\includegraphics[width = 0.4\linewidth]{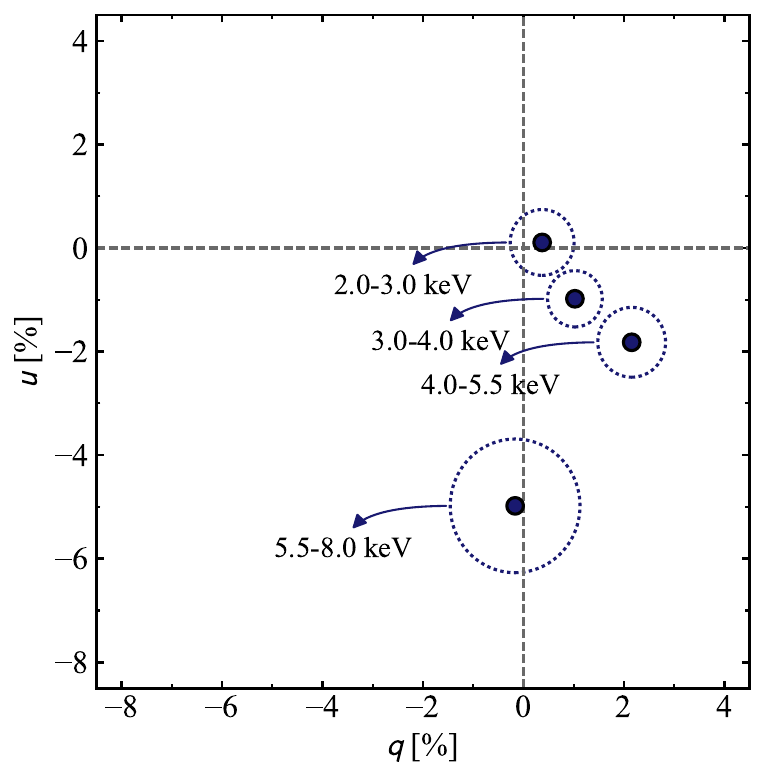}
\includegraphics[width = 0.4\linewidth]{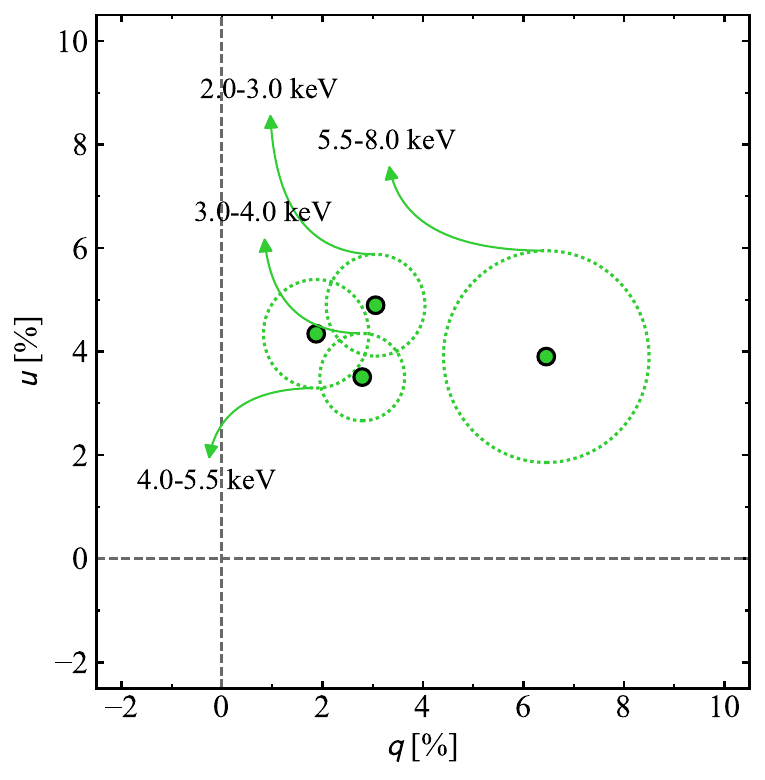}
\caption{Same as Fig.~\ref{fig:pdpa_ene_split}, but for the normalized Stokes parameters $q$ and $u$. Arrows only indicate the label of the energy band. Uncertainties are reported at $68\%$ CL.
}
\label{fig:stokes_ene}
\end{figure*}

However, the pre- and post-dip parts show an interesting and completely different behavior (see Table~\ref{tab:PCUBE}.) 
The energy-resolved PA and PD for these parts are presented in Fig.~\ref{fig:pdpa_ene_split}, and the corresponding Stokes $q$ and $u$ parameters are shown in Fig.~\ref{fig:stokes_ene}. 
We also additionally studied the dip core, one specific bin in Fig.~\ref{fig:lightcurve} at $t=6.15$~d that falls out of the patterns in Fig.~\ref{fig:lightcurve}a and Fig.~\ref{fig:lightcurve}b. This bin lies in the middle of the hardening in Fig.~\ref{fig:lightcurve}e and of the dip in Fig.~\ref{fig:lightcurve}f, which makes it particularly interesting. As mentioned in Table~\ref{tab:PCUBE}, the PD there exceeds $7\%$, and the PA is below $-50\degr$ with a significance of the detection higher than $99\%$.

The change in PD behavior is noticeable. 
In the pre-dip part, the PD strongly depended on energy and increased from an upper limit of $1.7\%$ at $2\sigma$ CL in the 2--3 keV bin to a strong detection of  $5.5\%$ in the 5.5--8 keV bin. 
In the post-dip part, however, the PD was almost constant with energy and averaged at about $5.3\%$, exceeding $7\%$ in the 5.5--8 keV bin. 
No significant PA dependence on energy is observed in either part of the observation, but the value of the angle shifts by almost 60\degr\ between the pre-dip and post-dip parts. 
This complete change in the behavior between the two parts of the observation is also visible in the Stokes $q-u$ plane (see Fig.~\ref{fig:stokes_ene}): The polarization vector changes in amplitude and direction. 

\begin{figure}
\centering
\includegraphics[width =0.9\linewidth]{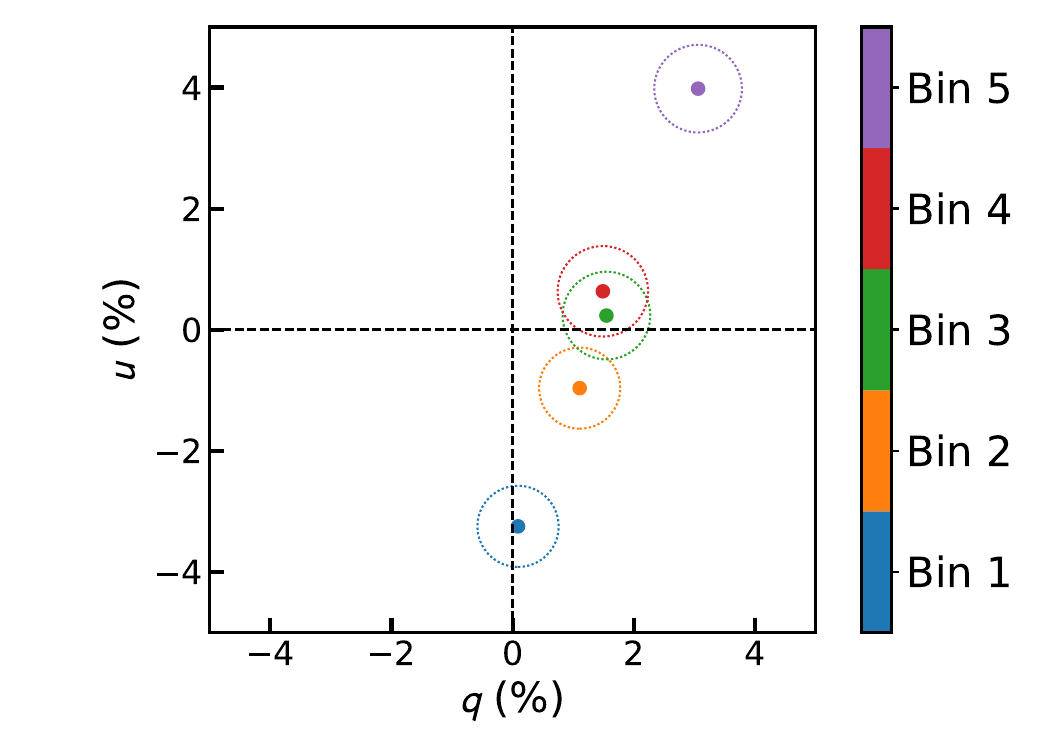}
\caption{Time dependence of the normalized Stokes parameters $q$ and $u$  obtained by the \texttt{pcube} algorithm using the data separation into five equal time bins. Uncertainties are reported at the $68\%$ CL.}
\label{fig:contour_5_bins}%not contour anymore, but for the flexibility let's keep it for now
\end{figure}

\begin{figure}
\centering
\includegraphics[width = 0.9\linewidth]{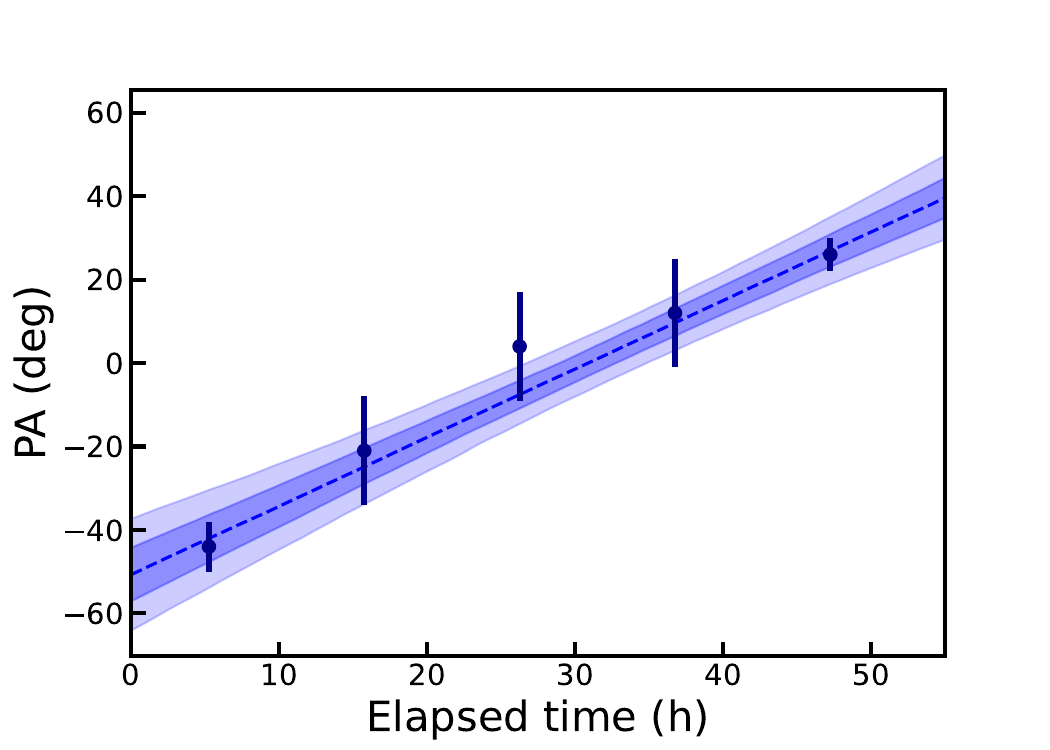}
\caption{Linear fit of the PA obtained for time bins~1--5 in Fig.\ref{fig:lightcurve}f. The plot illustrates the rotation of the polarization plane with time. 
Uncertainties are reported at the $68\%$ CL.
}
\label{fig:line}
\end{figure}

To study the separation of the observation in even more detail, we split it into five equal parts of 10.5~h each (bins~1--5 are defined with vertical dashed blue lines  in Fig.~\ref{fig:lightcurve}f). 
The results are shown in Fig.~\ref{fig:contour_5_bins} and Table~\ref{tab:PCUBE}. 
We performed a fit of the five PA values with a straight line (see Fig.~\ref{fig:line}) and confirmed the clear rotation of the PA with time with a slope of 1\fdg64 per hour and a total rotation angle exceeding 70\degr. 
We also note that the PD is relatively high at the beginning and end of the observation. 
We note here that the polarization in bins~2, 3, and 4 in Fig.~\ref{fig:contour_5_bins} is consistent with zero within the error bars at 99\% CL. 
Thus, as three individual measurements, they are not significant, but as a part of the trend, they are still meaningful. The important conclusion we make here is that the dip does not fully cause the variability in the polarimetric properties. As can be concluded from the data presented in Table~\ref{tab:PCUBE}, the drop in PD already occurred in bin 2, while the dip occurs around bin 4. The PA rotates throughout the whole observation. The anomalies in the PA and PD occur around the time corresponding to the dip in the light curve in Fig.~\ref{fig:lightcurve}, but the general trend for PA and PD is not directly correlated with the dip.

Last but not least, we attempted to divide the observation into two parts by the count rate: below and above 25 cnt\,s$^{-1}$. 
Data below the threshold fall into one of the many dips in the light curve. In the brighter state (above 25 cnt\,s$^{-1}$), we saw a PD of $\approx1.6\%$ and PA of $\approx2\degr$, which is similar to that shown in Fig.~\ref{fig:pdpa_ene_ave} and in Table~\ref{tab:PCUBE} for the overall polarization. The dimmer state (below 25 cnt\,s$^{-1}$) showed $\mbox{PD}\approx1.3\%$ and $\mbox{PA}\approx-36\degr$, which is in between the wide dip state of the previous analysis (between MJD$-$60230=6.05 and 6.30, as shown in Fig.~\ref{fig:lightcurve}f) and the overall polarization. Thus, adding the data obtained from the times corresponding to all the minor dips smoothed the difference between the anomalous polarimetric behavior during the dip and the general trend that we observed. It shifted the dimmer state PA and PD closer to the overall states than those of the wide dip. The polarimetric properties of the minor dips are then different from those of the wide dip, which means that the nature of the wide dip is different from the nature of all other dips in the light curve we observed.

\subsection{Spectroscopic analysis}
\subsubsection{\xmm}

\begin{figure}
\centering
\includegraphics[width = 0.9\linewidth]{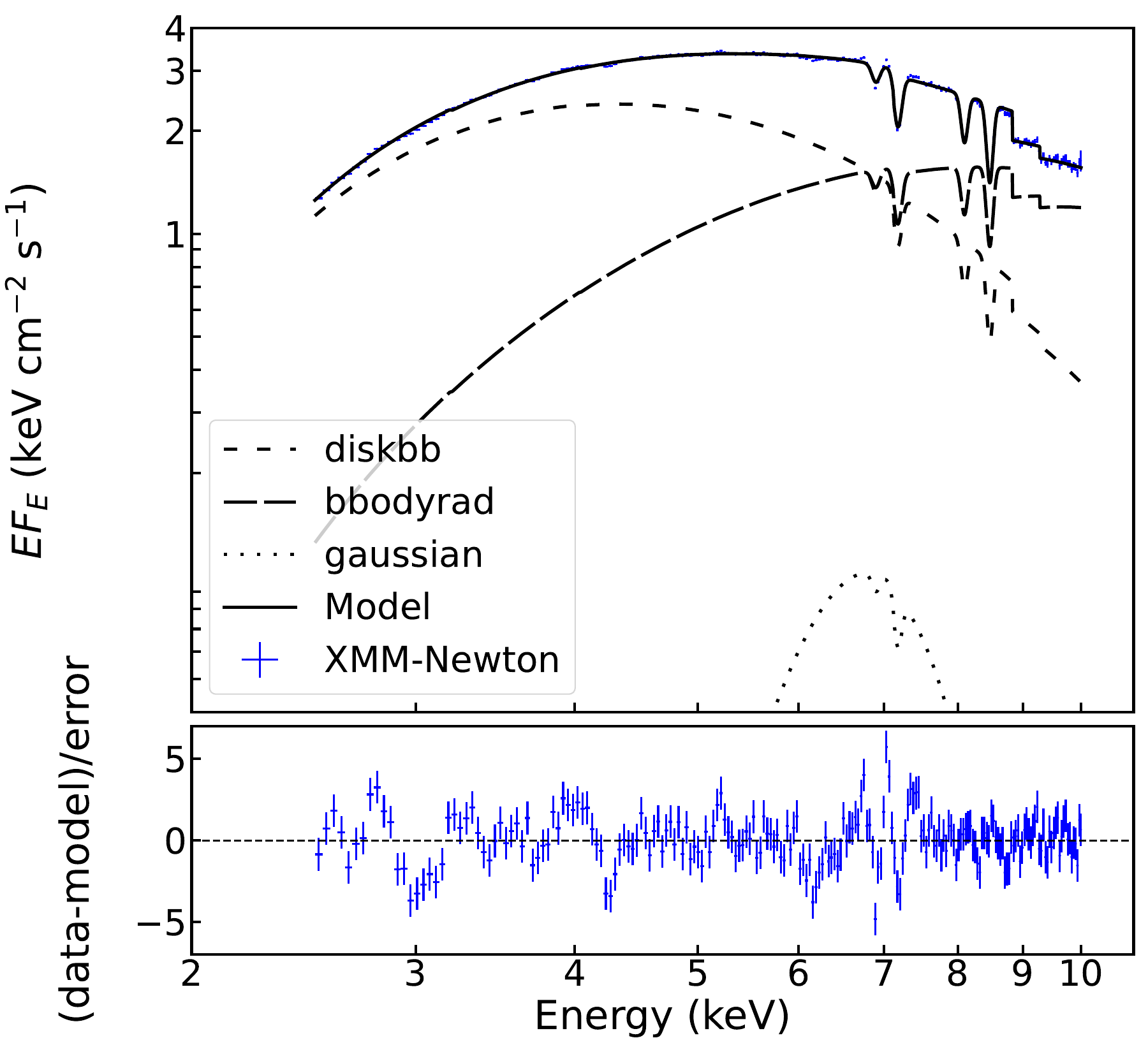}
\caption{Spectral energy distribution of \source in $EF_E$ representation. 
The blue crosses show the data from \xmm. 
The different spectral model components are reported as black lines for \texttt{diskbb} (short dashed), \texttt{bbodyrad} (long dashed), and \texttt{gaussian}  (dotted). 
The total model is shown with a solid black line. 
The bottom panel shows the residuals between the data and the best fit. 
The data are rebinned only for plotting purposes.}
\label{fig:xmm_spec}
\end{figure}

\begin{table} % [h!]
\centering
\caption{Best-fit parameters of \source spectrum from \xmm data.} 
%\scriptsize
\footnotesize
\begin{tabular}{ll c }
\hline
\hline
\multicolumn{2}{c}{Parameter} & Value  \\
 %        & & &       \\  
\hline 
{\tt tbabs}  & $N_{\rm H}$ ($10^{22}$  cm$^{-2}$)& $5.94^{+0.06}_{-0.07}   $ \\
{\tt diskbb} &  $kT$  (keV) & $1.39^{+0.12}_{-0.06}$ \\
&  norm & $130^{+20}_{-30}$ \\
{\tt bbodyrad} & $kT$ (keV) & $2.0^{+0.3}_{-0.1}$ \\ 
& norm & $20.08 \pm 0.03$  \\ 
& $R_{\rm bb}$ (km) & $3.13 \pm 0.02$ \\
{\tt gaussian} & $E$ (keV) & $6.6$ (frozen)\\ 
& $\sigma$ (keV) & $0.8$ (frozen)\\
& norm & $0.0048^{+0.0009}_{-0.0007}$\\
{\tt gabs} & $E$ (keV) & $8.476^{+0.005}_{-0.009}$ \\
& $\sigma$ (keV) & $0.05$ (frozen) \\
& depth (keV) & $0.067\pm0.004$ \\
%{\tt gabs} & $E$ (keV) & $4.29$ (frozen) \\
%& $\sigma$ (keV) & $0.05$ (frozen) \\
%& depth (keV) & 1e-6 (frozen)\\
{\tt gabs} & $E$ (keV) & $8.10\pm0.01$   \\
& $\sigma$ (keV) & $0.05$ (frozen) \\
& depth (keV) & $0.040\pm0.003$ \\
{\tt gabs} & $E$ (keV) & $7.180^{+0.006}_{-0.004}$ \\
& $\sigma$ (keV) & $0.05$ (frozen) \\
& depth (keV) & $0.040\pm0.001$ \\
{\tt gabs} & $E$ (keV) & $6.90\pm0.01$ \\
& $\sigma$ (keV) & $0.050$ (frozen) \\
& depth (keV) & $0.0150^{+0.0017}_{-0.0013}$ \\
{\tt edge} & $E$ (keV) & $8.83$ (frozen) \\
& $\tau$ & $0.20\pm0.01$\\ 
{\tt edge} & $E$ (keV) & $9.28$ (frozen) \\
& $\tau$ & $0.08\pm0.01$\\  
 \hline
&  $\chi^2$/d.o.f.  &  1504/1483
\\ 
\hline
\end{tabular}
\tablefoot{The errors are at 68\% CL. $R_{\rm bb}$ is calculated from the normalization of the \texttt{bbodyrad} component, assuming a distance to the source of 7 kpc. }  
\label{tab:XMMspectrum}
\end{table}   

In the \xmm data, below 2 keV, the count rate drops significantly due to the high interstellar absorption. 
We observe a significant excess below 2 keV with any continuum model combination. 
High residuals are also seen near the instrumental edges. 
In combination with our inability to subtract the background, this led us to discard these low-energy data. 
After all, they do not affect the modeling at higher energies.

Thus, we modeled the spectrum of the \xmm observation in the 2--10 keV range. 
As we traditionally expect two continuum components to be present in the spectrum of a WMNS with some additional reflection and absorption features, we performed a fit with a
\texttt{tbabs*(diskbb+bbodyrad+gaussian)} model modified by four Gaussian absorption lines, \texttt{gabs}, and two interstellar dust absorption edges, \texttt{edge}. 
The results are shown in Table \ref{tab:XMMspectrum} and in Fig.~\ref{fig:xmm_spec}. 
To properly resolve the absorption features in the 6--9 keV energy range, we fixed the parameters of the \texttt{gaussian} corresponding to the broad emission line. 

The resulting fit illustrates that the spectrum of \source is dominated by the disk.
The harder component is modeled with the blackbody \citep[as suggested in][]{2003A&A...406..221S, DT12, Revnivtsev2013} and dominates at energies higher than 7 keV. 
Four strong absorption lines at 6.90, 7.18, 8.10, and 8.48~keV are also visible. 
These lines are identified with the Fe XXV He$\alpha$ resonant line at 6.700 keV, Fe XXVI Ly$\alpha$ doublet at 6.966 keV, Ni XXVII He$\alpha$ resonant line at 7.806 keV, and Ni XXVIII Ly$\alpha$ doublet at 8.092 keV, and they are well known in the source \citep[see, e.g.,][]{2020MNRAS.497.4970T}; three other known lines are not resolved in our fit. 
Compared with the previous results, the energies of the lines we obtain are higher than in \citet{DT12} and correspond to a very high blueshift, $z=-0.03$ for the Fe lines and even higher for the Ni lines (although the latter are known to be blended with Fe He$\beta$ lines) and might come from the full science buffer of the \xmm during our observation. It is known that in this source, the wind is complex, very structured, and highly variable, and we therefore avoid interpreting this nominal blueshift without a much more detailed analysis of the absorption features, which is well beyond the scope of this paper. However, we do not expect this issue to significantly affect the continuum parameters in which we are mostly interested.
We also note the anomalously high equivalent hydrogen column density parameter $N_{\rm H}$ of the absorption component.

Additionally, we tried to model the \xmm spectrum with only one continuum component. 
That is, we removed the \texttt{bbodyrad} component, and fit the whole continuum using a single \texttt{diskbb} component with high $kT_{\rm dbb} \approx 1.9$~keV, resulting in a bad $\chi^2$/d.o.f.= 2058/1485 corresponding to a null-hypothesis probability $<10^{-3}$\% with respect to the  34\% achieved from the fit reported in Table~\ref{tab:XMMspectrum}. 

\subsubsection{IXPE}

\begin{figure}
\centering
\includegraphics[width = 0.85\linewidth]{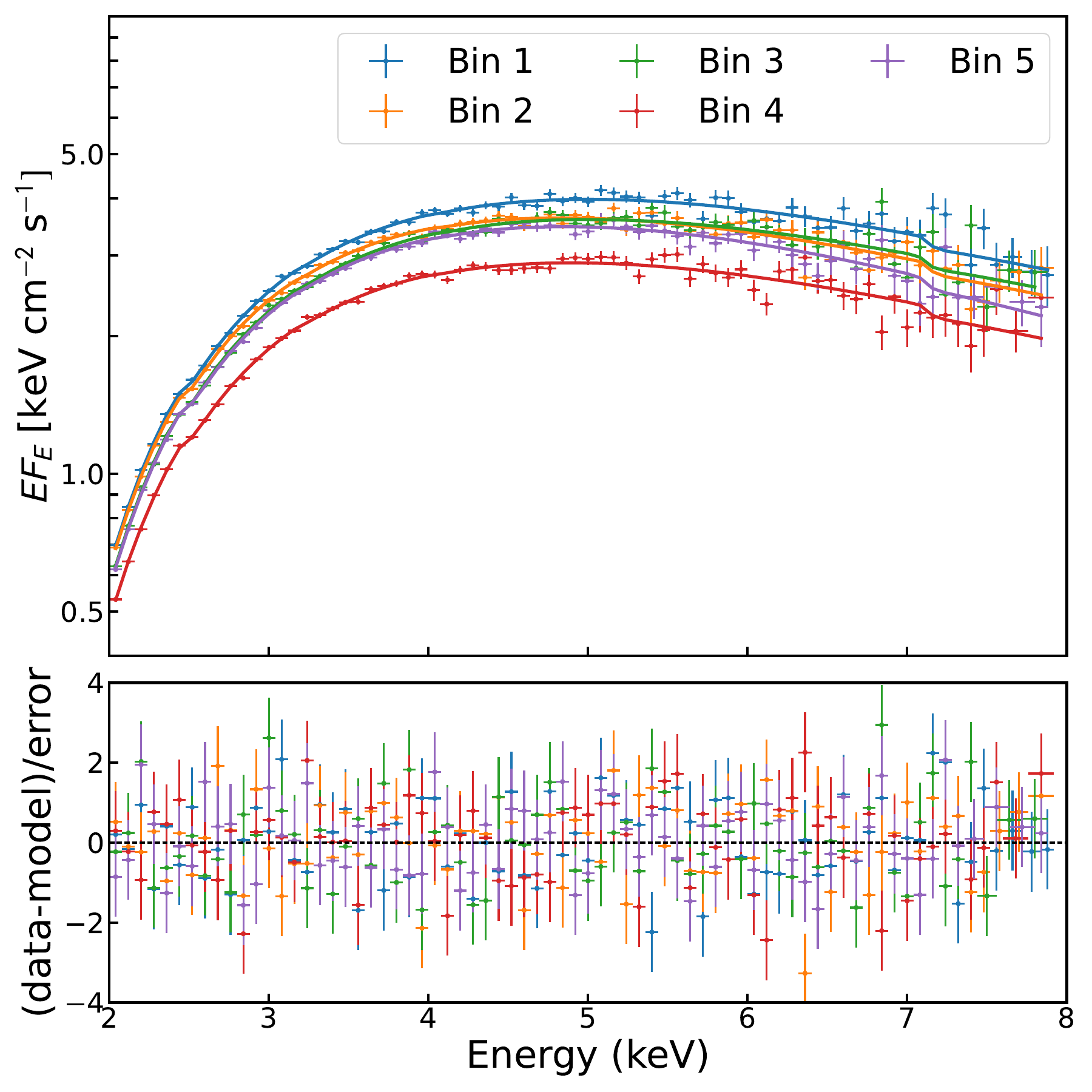}
\caption{Spectral energy distribution of the source in $EF_E$ representation during five equal parts of the observation. 
Here, only \ixpe DU1 is reported as an example, and the data are rebinned for plotting purposes. }
\label{fig:5_spectra}
\end{figure}

\begin{table*}
\centering
\caption{Best-fit parameters of \source spectrum from \ixpe data divided into five equal time bins.} 
\begin{tabular}{ccccccc}
\hline
\hline
Component & {Parameter (units)} & Bin~1 & Bin~2 & Bin~3 & Bin~4 & Bin~5 \\ \hline
\texttt{tbabs} & $N_{\rm H}$ ($10^{22}$ cm$^{-2}$) & 5.9 (frozen) & 5.9 (frozen) & 5.9 (frozen) & 5.9 (frozen) & 5.9 (frozen) \\
\texttt{diskbb} & $kT_{\rm in}$ (keV) & 1.49$^{+0.03}_{-0.05}$ & 1.44$^{+0.04}_{-0.03}$ & 1.45$\pm0.04$ & 1.57$_{-0.07}^{+0.06}$ & 1.55$\pm$0.06\\
& \texttt{norm} & 131$^{+14}_{-8}$ & 142$^{+13}_{-12}$ & 130$_{-11}^{+10}$ & 79$_{-8}^{+10}$ & 
102$_{-7}^{+10}$\\
\texttt{bbodyrad} & $kT$ (keV) & 2.0 (frozen) & 2.0 (frozen) & 2.0 (frozen) & 2.0 (frozen) & 2.0 (frozen) \\
& \texttt{norm} & 14$\pm4$ & 14$^{+2}_{-5}$ & 14$\pm3$ & 10$\pm4$ & 8$^{+4}_{-5}$\\
& $R_{\rm bb}$ (km) & 2.6$\pm 0.4$ & 2.6$^{+0.2}_{-0.5}$ & $2.6\pm0.3$ & $2.2\pm0.4$ & $2.0^{+0.5}_{-0.6}$\\
\texttt{constant} & DU1 & 1 (frozen) & 1 (frozen) & 1 (frozen) & 1 (frozen)& 1 (frozen) \\
& DU2 & $0.951\pm0.004$ & $0.956\pm0.004$ & $0.948\pm0.004$ & $0.951\pm0.004$ & $0.950\pm0.006$\\
& DU3 & $0.908\pm0.003$ & $0.910\pm0.004$ & $0.904\pm0.004$ & $0.912\pm0.004$ & $0.911\pm0.005$ \\
\hline
& $\chi^2$/dof & 392/420 & 415/416 & 389/411 & 420/411 & 403/409 \\
\hline
\multicolumn{2}{c}{Flux$_{\textrm{2-8\,keV}}$ ($10^{-9}$ \flux)} & 6.42 & 5.98 & 5.77 & 4.96 & 5.65 \\
\multicolumn{2}{c}{Flux$_{\texttt{diskbb}}$/Flux$_{\textrm{2-8\,keV}}$} & 0.82 & 0.83 & 0.81 & 0.81 & 0.91 \\
\multicolumn{2}{c}{Flux$_{\texttt{bbodyrad}}$/Flux$_{\textrm{2-8\,keV}}$} & 0.18 & 0.17 & 0.19 & 0.19 & 0.09 \\ 
\hline
\end{tabular}
\tablefoot{The errors are at the 68\% CL. A gain slope of $\approx$0.96 and gain offsets $\approx$2--80 eV for the three DUs were needed to obtain a good reduced $\chi^2$. $R_{\rm bb}$ is calculated from the normalization of the \texttt{bbodyrad} component, assuming the distance to the source of 7 kpc. } 
\label{tab:IXPEspec}
\end{table*}

As the HID obtained from the \ixpe data in Fig.~\ref{fig:HID} has rather large uncertainties and was obtained over a small energy range, it is not the most informative representation of the spectral variation in the data with time. 
We used \textsc{xspec} to fit the $I$ spectrum obtained by \ixpe. 
For this study, we chose the same five equal time bins as for the \texttt{pcube} analysis presented in Fig.~\ref{fig:contour_5_bins}. 
All five spectra were fit with the same model \texttt{tbabs*(diskbb+bbodyrad)}. 
Considering the spectral capabilities of \ixpe and the small effective area above 6 keV, we did not attempt to add the \texttt{gabs} or \texttt{gaussian} components that were prominent in the \xmm spectrum. We multiplied the model by the \texttt{constant} component to scale the data from the three detectors of \ixpe. 
We also fixed the value of the equivalent hydrogen column density $N_{\rm H}$, as \ixpe does not have the capability to estimate it, and the temperature of the \texttt{bbodyrad} component to the best-fit values obtained with the \xmm. We applied the weighted analysis introduced in \citet{DiMarco2023}. 
The results are presented in Table \ref{tab:IXPEspec} and Fig.~\ref{fig:5_spectra}. 

We note the lack of a change in the spectral parameters with time. 
The temperature of the disk does not vary with time, and only the normalization of the continuum components decreases toward the end of the observation, which could be consistent with increased absorption or scattering in the disk wind during the last part of the observation. 
In bin~5 of the observation, the contribution of the blackbody component also drops by a factor of two (however, the normalizations of all the blackbody components are compatible with each other within a 1 $\sigma$ error). 
Bin~4 almost coincides with the broad dip and has a lower flux, and this is shown in Fig.~\ref{fig:5_spectra}.

\begin{figure*}
\centering
\includegraphics[width = 0.9\linewidth]{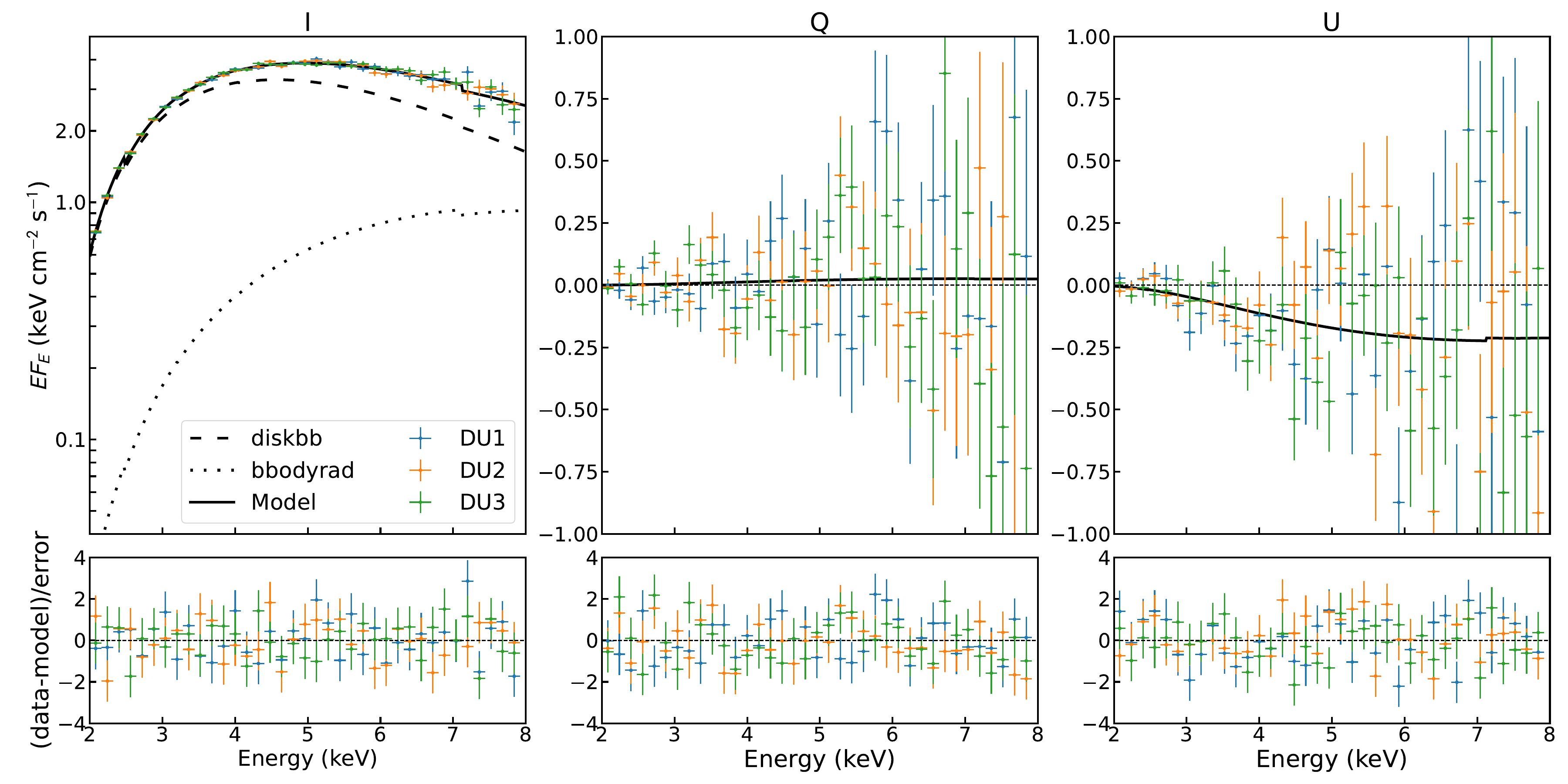}
\caption{Spectral energy distribution of \source in $EF_E$ representation as observed by \ixpe in bin~1. 
The left, middle, and right panels show the Stokes parameters $I$, $Q$, and $U$, respectively.  
The fit is performed in the 2--8 keV energy band using the three \ixpe detectors and applying the model \texttt{tbabs*pollin*(diskbb+bbodyrad)}. 
The total model is shown with the solid black line, and \texttt{diskbb} and \texttt{bbodyrad} are shown with the dashed and dotted lines, respectively. 
The lower subpanels show the residuals between the data and the best fit. 
The data are rebinned only for plotting purposes.
}
\label{fig:IQU}
\end{figure*}

\begin{table*}
\centering
\caption{Best-fit parameters of the spectropolarimetric model for \source. } 
\begin{tabular}{cccccc}
\hline
\hline
Component & {Parameter (units)} & Bin~1, \texttt{polconst} & Bin~1, \texttt{pollin} & Bin~2--4 & Bin~5 \\ \hline
\texttt{tbabs} & $N_{\rm H}$ ($10^{22}$ cm$^{-2}$) & 5.9 (frozen) & 5.9 (frozen) & 5.9 (frozen) & 5.9 (frozen)\\
\texttt{diskbb} & $kT_{\rm in}$ (keV) & $1.51_{-0.04}^{+0.05}$ & $1.51_{-0.04}^{+0.05}$ & $1.50_{-0.03}^{+0.02}$ & $1.58_{-0.06}^{+0.05}$ \\
 & norm & $120\pm 11$ & $120 \pm 11$ & $106_{-5}^{+6}$ & $92 \pm 10$ \\
\texttt{bbodyrad} & $kT$ (keV) & 2.0 (frozen) & 2.0 (frozen) & 2.0 (frozen) & 2.0 (frozen) \\
 & norm & $13_{-3}^{+6}$ & $13_{-3}^{+4}$ & $12 \pm 2$ & <14 \\
 & $R_{\rm bb}$ (km) & $2.5_{-0.3}^{+0.6}$ & $2.5_{-0.3}^{+0.4}$ & $2.4\pm0.2$& <2.6 \\
\texttt{polconst} & $A$ (\%) & $3.0\pm0.5$ & -- & $1.2\pm0.3$ & $4.5\pm0.5$ \\
 & $\psi$ (deg) & $-42\pm5$ & -- & $7\pm7$ & $32\pm3$ \\
\texttt{pollin} & $A_1$ (\%) & -- & $-0.75 \pm 1.4$ & -- & --\\
 & $A_{\rm slope}$ (\%\,keV$^{-1}$) & -- & $1.3\pm0.5$ & -- & --\\
 & $\psi_1$ (deg) & -- & $-42\pm4$ & -- & --\\
 & $\psi_{\rm slope}$ (deg\,keV$^{-1}$)  & -- & 0 (frozen) & -- & -- \\
\texttt{constant} & DU1 & 1 (frozen) & 1 (frozen) & 1 (frozen) & 1 (frozen) \\
 & DU2 & $0.953\pm0.004$ & $0.953\pm0.004$ & $0.956\pm0.002$ & $0.950\pm0.004$ \\
 & DU3 & $0.910\pm0.004$ & $0.910\pm0.004$ & $0.910\pm0.002$ & $0.909\pm0.004$ \\
\hline
 & $\chi^2$/d.o.f. & 634/653 & 626/652 & 600/653 & 630/653\\ \hline
\end{tabular}
\label{tab:spectropol}
\tablefoot{\ixpe $I$, $Q$, and $U$ spectra are divided into five equal time bins and regrouped based on their spectral and polarimetric behavior.
The errors are at the 68\% CL. A gain slope of $\approx$0.97 and gain offsets $\approx$130--200 eV for the three DUs were needed to obtain a good reduced $\chi^2$. $R_{\rm bb}$ is calculated from the normalization of the \texttt{bbodyrad} component, assuming the distance to the source of 7 kpc.} 
\end{table*}

\subsection{Spectropolarimetric analysis}

For the spectropolarimetric analysis, we used the results of the model-independent polarimetric analysis and the spectroscopic analysis as a driver. 
As we learned that both spectral and polarimetric properties are rather similar in bins 2--4 (see Table~\ref{tab:PCUBE} and Table~\ref{tab:IXPEspec}), we combined these three in one larger time bin. Our aim was to obtain a stronger detection in this part of the observation, as in the polarimetric analysis we obtained the results that are only meaningful as a part of the trend for bins 2--4. We studied bins~1 and~5 separately. 
Instead of studying the polarization of each spectral component independently, we applied the \texttt{polconst} model to the entire continuum, as the data quality did not allow us to determine the polarization of the two spectral components. For bin~1, we also attempted to apply the \texttt{polllin} polarimetric model, as we previously saw the dependence of PD on energy at the beginning of the observation (see Fig.~\ref{fig:pdpa_ene_split}, left). 
We again fixed the hydrogen column density $N_{\rm H}$ and the blackbody temperature. 
The results are presented in Table \ref{tab:spectropol}, and the fit of three Stokes parameters, $I, Q$, and $U$, performed for the bin~1 data with the \texttt{pollin} model is shown in Fig.~\ref{fig:IQU}. The distribution of the residuals for all the three parameters illustrates the quality of the fit, which was found to be good.

We note that for bins 1 and 5, the spectral and polarimetric properties obtained in the spectropolarimetric analysis agree with the previous results presented in Tables~\ref{tab:PCUBE} and \ref{tab:IXPEspec}. As we do not have strong detection for the polarimetric properties of bins~2--4, we decided not to present the spectropolarimetric analysis of these bins independently in Table~\ref{tab:spectropol}, as this does not improve the quality of our results. 
In bin~1, due to the strong dependency of PD on energy, we applied the \texttt{pollin} model with a constant PA, which gives an improvement by $\Delta\chi^2=9$ compared to the fit with the \texttt{polconst} model (which has one parameter less). 
This improvement corresponds to an F-test probability of $\approx3\times10^{-3}$. 

\section{Discussion}
\label{sec:discussion}
\subsection{Correlation of the polarimetric behavior with the dip in the light curve}

The data analysis presented in Sect.~\ref{sec:analysis} provided us with several significant insights that require interpretation and further reflection. 
To begin with, we observed significant and anomalous variation in the light curve: the broad dip closer to the end of the observation is  not correlated with the regular dips known in the source, as the folding of the \maxi data using the formulae from \citet{2014A&A...561A..99I} suggests that our observation was made one week before the regular dip. 
During the dip, we also observed a slight increase in the hardness ratio. It would be natural to expect that polarimetric properties, as hardness and the count rate, have some variations around 6.1-6.3 [MJD--60230].
However, the behavior of the polarimetric properties is not aligned with this variation time-wise: There is a polarimetric anomaly during the dip core state, but the overall trend in PA and PD is not correlated with the dip. PA increases slowly throughout the whole observation, and PD decreases already in bin 2 and then increases in bin 5 without a direct correlation to the dip in the light curve. 
The inconsistency in the behavior of the spectral and polarimetric properties invites us to search for an explanation for the PA and PD variation outside the regular hardness/intensity changes. 

The results of the previous \ixpe observations of the X-ray binaries showed that the concept of a rapid and significant change in the PA is not new to this class of objects. 
The high luminosity of \source allowed us to separate the observation into smaller parts and see the slow monotonous rotation of the polarization plane, but the results presented in Fig.~\ref{fig:pdpa_ene_split} are similar to several results that were obtained previously. 
It is tempting to assume that similar mechanisms might cause the polarimetric behavior of \source and that a slow rotation similar to the one presented in Fig.~\ref{fig:contour_5_bins} has not been observed previously just due to the lack of statistics or a different approach used in the data analysis. 
For instance, in \mbox{Sco X-1}, \cite{LaMonaca2024} reported that the rotation of the PA with respect to the previous observations \citep{1979ApJ...232L.107L} and the jet position angle was around $54\degr$, and the authors explained it with the precession of the jet or with the change in the corona geometry with the transition between the states. 
However, no sign of an X-ray color variability in \source could be associated with these state changes.  
In the case of \mbox{Cir X-1}  \citep{Rankin2024}, the PA changed by $60\degr$ between the two parts of a single observation, but in this case, a clear sign of a corresponding state transition was observed. 
The difference of $60\degr$ between the PA of the hard and soft components of the observation was explained by a switch from the BL to the SL in a tilted NS, as \mbox{Cir X-1} is a young system \citep{2013ApJ...779..171H}. 
\source, on the other hand, is an older stable system (as is suggested by the companion being a late-evolved K5 III giant) with a high accretion rate; the chances for the rotation axis of the NS to be significantly misaligned relative to the orbital axis are smaller. 
%Misalignment seems to be an unlikely explanation for such a system.
The results are in many ways also very similar to the polarization that is commonly seen in black hole X-ray binaries (see, e.g., Fig.\,S6 in \citet{Krawczynski2022} in comparison with Fig.~\ref{fig:pdpa_ene_split}). 
As for \mbox{Cyg~X-1}, in \source we see the PD during a higher flux rising with energy, while the PD at a (slightly) lower flux is constant. 
The PA is constant with energy in both cases, but the change in the PA by $60\degr$ between the two cases presented in Fig.~\ref{fig:pdpa_ene_split} is significantly different from that in black holes. 
The rotation of the PA with time was seen recently in the blazar \mbox{Mrk~421} \citep{2023NatAs...7.1245D}. They explained it by the helical magnetic structure in the jet illuminated in the X-rays by a localized shock propagating along this helix. Unfortunately, jets have only been observed in \source in the hard state \citep{2018ApJ...861...26A}, and we observed the source in the soft state. It is therefore challenging to imagine that a similar geometry and physics cause the PA rotation in \source as in the \mbox{Mrk 421}. 
In X-ray pulsars, the rotation of the PA with the pulsar phase is routinely observed \citep{Doroshenko22,Doroshenko23,Tsygankov22,Tsygankov2023}, but \source is not a pulsar.
We conclude that we need to search for a new mechanism that causes the polarimetric behavior of \source. 

\subsection{Additional polarimetric component}

\begin{figure}
\centering
\includegraphics[width =\columnwidth]{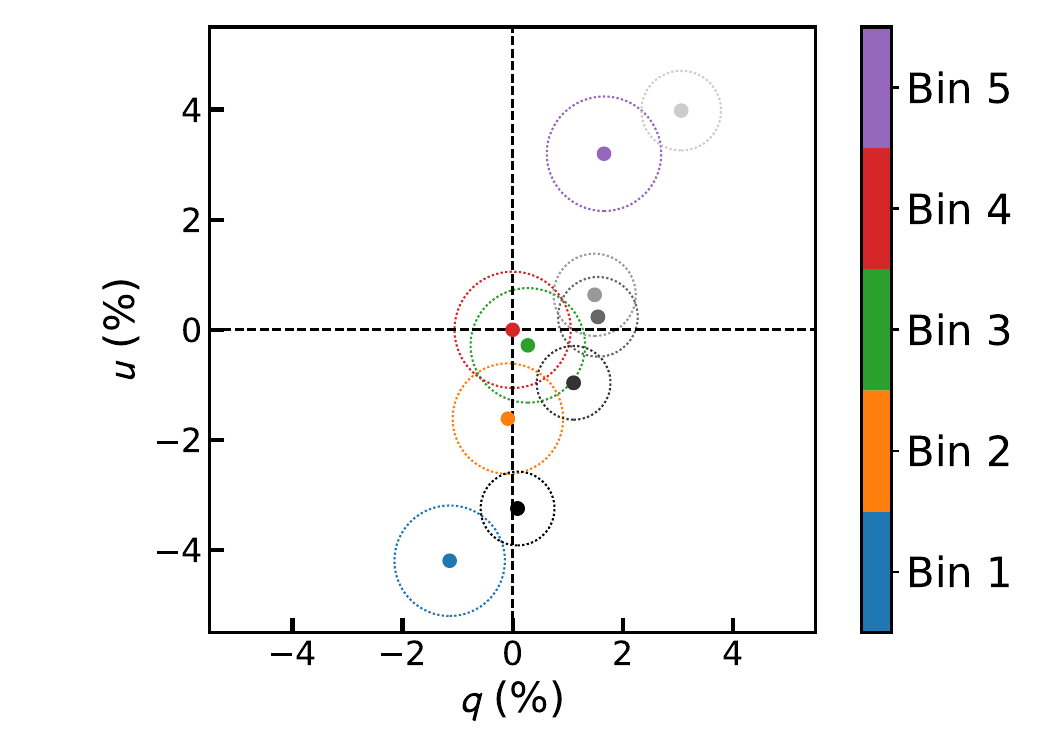}
\caption{Same as Fig.~\ref{fig:contour_5_bins}, but the (absolute) Stokes parameters corresponding to bin~4 are subtracted. 
The gray points are the same as in Fig.~\ref{fig:contour_5_bins}.  }
\label{fig:qu_fivebins}
\end{figure}

%The evolution of normalized Stokes parameters $q$ and $u$, as depicted in Fig.~\ref{fig:contour_5_bins}, offers an additional perspective on the puzzle of polarization variability of the source.
Fig.~\ref{fig:contour_5_bins}
%Notably, 
showed that the source 
%does not simply show a gradual rotation of the PA but a 
evolves in a straight line (within the uncertainties) in the $q-u$ plane shifted from the origin. 
Considering this, we tried to decompose the polarized flux into different components with different polarization orientations. 
It would be problematic to interpret the emission in bins~1 and ~5 as coming from two distinct components that differ in PA by 70\degr, with the rest representing a transitional phase, because a transition like this is unlikely to result in a straight-line trajectory in the $q-u$ plane, particularly given that the flux during bin~4 is lower than during bins~1 and~5. 
We therefore suggest alternatively that the polarization nature in bins~1 and~5 is predominantly the same, and that the misaligned polarization component is observed in between.
A similar model that involves a constant polarized component has previously been proposed to interpret X-ray polarization data from the bright X-ray pulsars \mbox{LS~V~+44~17}/ \mbox{RX~J0440.9+4431} 
\citep{Doroshenko23} and Swift~J0243.6+6124 \citep{Poutanen24}. 

We define a constant component corresponding to bin~4 (a period nearly overlapping with the dip, as shown in Fig.~\ref{fig:lightcurve}f).
Thus, we subtract absolute Stokes parameters corresponding to bin~4 from the absolute Stokes parameters of each time bin. 
The resulting normalized Stokes parameters are shown in  Fig.~\ref{fig:qu_fivebins} with colored points, and the originally observed values are shown with gray points.
Their positions are still consistent with a straight line that is now located near the origin. 
In this scenario, the polarization vectors during bins~1 and~5 are essentially perpendicular to each other (since they lie on the same line in the $q-u$ plane at different sides around the origin), 
and the radiation observed in bins~2 and~3 appears to be statistically consistent with being unpolarized. 

The picture we see after the subtraction is physically meaningful. A difference of 90\degr\ in PAs of bins~1 and~5 could be attributed to changes in the geometry of the scattering or reflecting medium, changes in the scattering optical depth, and the relative brightness changes in the accretion disk, the BL, and the SL. The PA of the permanent component of the polarization (in our assumption, similar to that of bin 4) differs from the PA of the varying component (which we see after the subtraction in Fig.~\ref{fig:qu_fivebins}) by $\approx30\degr$, as clearly visible in the $q-u$ plane. However, as technically, we could have associated the permanent component with any other bin, it is impossible to constrain the angle. The main conclusion here is that this angle is nonzero. To provide this misalignment of the PA between the components associated with the different emission regions within the source, a misalignment in the geometry of the system should be introduced.

\subsection{Influence of the variations in the wind}

One other possible explanation for the rotation of the polarization plane with time could be the impact of the orbital movement on the polarization arising from the scattering in the accretion bulge. 
With the high accretion rate known in \source, some asymmetry in the accretion disk surrounding the NS can be expected. 
However, \source is recognized for having the largest disk in terms of gravitational radius of all observed LMXBs \citep{2020MNRAS.497.4970T}. 
If a bulge exists, it is therefore probably quite distant from the source. 
Moreover, we observed a rotation of the polarization plane by $\approx70\degr$ in the two days of observation. 
The orbital period of \source is 24.5~d. 
The unknown mechanism that causes the rotation of the polarization plane would need to operate on a shorter timescale. 

\source is known to have a strong and rapidly changing wind, and evidence for the absorption in the wind comes from the \xmm observation. 
Scattering in the equatorial wind can produce different polarization patterns in theory, and with this mechanism and with the high inclination of the source, relatively high polarization can be produced easily (see \citealt{SuTit1985, Tomaru2024}; Nitindala et al., in prep.). 
The nature of the polarization variability could be that the observed emission is scattered in the different parts of the wind above the accretion disk. 
As different absorption features are assumed to come from several distinct parts of the wind with various physical, optical, and geometrical properties, we expect significant differences in the polarimetric behavior. 
If the line of sight passes close to the edge of the wind, then small variations in the wind opening angle would lead to large variations in the contribution of the scattered component and the polarization degree. 
%One way to produce such a rapid change in the wind scattering is to observe the source at an inclination from the opening angle of the wind. As the opening angle varies with time, the observer's line of sight would pass either through the wind or above the wind, which could result in a completely different spectro-polarimetric picture. 
We hope to study the correlation between the behavior of the wind and polarimetric properties in the future with simultaneous  and \ixpe observations.

\subsection{Spectral features}

%Comparison of the spectrum observed with \xmm with the previous campaigns on \source reported in \citet{DT12} and \citet{2023MNRAS.522.3367S} allowed us to assume that the state of the source was somewhere between the one of Obs. 6 and Obs. 7 in \citet{DT12}. 
%However, these two observations are separated by 11 days, and the fast variability of the source suggests that \source could have made a full route over the HID within these 11 days. It is therefore hard to predict the spectrum of the source during the \ixpe observation from the results presented in \citet{DT12} and other archival observations. 
 
All the studies we performed on the spectra of \source confirmed the original results from the HID: according to (however limited) \ixpe data, there are no significant changes in the spectral behavior with time during the \ixpe observation. The study of the radius of the blackbody component $R_{\rm bb}$ showed no variation in the geometry throughout the observation, with only a slight decrease in the last two bins. In bin~5, the contribution of this component to the total flux also dropped. This adds a challenge to the data interpretation. When we assume that in bin~5, the polarization is produced by the dominating softer disk component, we need to find a way for the disk emission to be polarized at the detected level.
The PD coming from the disk is estimated at up to $4\%$ for an inclination of 80\degr\ and up to only $1.5\%$ for an inclination of 60\degr\ \citep{2022A&A...660A..25L} assuming a semi-infinite atmosphere dominated by electron scattering \citep{Chandrasekhar1960}.
This means that some scattering in the wind above the disk might need to be introduced. 
The spectrum contains strong absorption features and \source is known to have a strong wind with several absorbing areas. We therefore assume that the wind adds to the polarization of this component. For the harder component, it is more likely to be produced by the BL than the SL. The SL can produce PD up to 1.5\% (Bobrikova et al., in prep), which is not sufficient to explain the high PD we see for higher energies (see Fig.~\ref{fig:pdpa_ene_split}). We also expect the polarization from the SL to be orthogonal to the polarization from the disk. We do not see this behavior: all the changes in PA with energy in Fig.~\ref{fig:pdpa_ene_split} and especially in Fig.~\ref{fig:pdpa_ene_ave} come from the averaging over time.

We also explored the possibility that the two continua components (\texttt{diskbb} and  \texttt{bbodyrad}) each have different polarization properties by fitting the data with the \textsc{xspec} model \texttt{tbabs*(polconst*diskbb+polconst*bbodyrad)*const}. 
We only attempted to apply this model to bin~1  because in this alone, the PD strongly depends on energy, so that it would be easier to resolve the two components. 
The fit suggested that the \texttt{diskbb} component is almost unpolarized (no significant polarization with the PD$<$2.9\% is obtained at 99\% CL), and the \texttt{bbodyrad} component has a PD of 10\% and a PA of $-42\degr$. 
However interesting, this suggestion makes it even harder to understand the relatively high polarization in bin~5, where the disk completely dominates the spectrum. 
The results of the observation are promising and exciting. They require further study supported by the modeling of the emission from different components and the emission that is scattered in the wind above the disk. 

\section{Summary}
\label{sec:summary}

We reported the highly significant detection of polarization from \source obtained by \ixpe. 
We performed the polarimetric, spectroscopic, and spectropolarimetric analysis of the \ixpe data. 
The two main components of the spectrum, the soft disk emission and the harder Comptonized component, were described by the \texttt{diskbb} and \texttt{bbodyrad} models, respectively. 
The spectral analysis was supported by a nonsimultaneous observation performed by \xmm. 
Due to the high energy resolution of \xmm, we added the \texttt{gaussian} component to describe the broad emission line corresponding to the reflection of the NS surface emission from the disk, four Gaussian absorption lines, \texttt{gabs}, and two edges, \texttt{edge}.  
For the different parts of the observation, we used the \texttt{polconst} and \texttt{pollin} models for the polarimetric properties. 
The overall PD of the \ixpe data was measured at $1.4\%$ at >$5\sigma$ CL, and the PA at $-2\degr$. 

From the in-depth study of the time variability of the polarimetric properties, we reported the strong rotation of the PA by $70\degr$ during the two observation days. 
The PD varied from $3.2\%$ to the nondetectable and then up to $5.0\%$. 
Moreover, we reported the change in the dependence of the PD on energy. At the beginning of the observation, we detected polarization that depended linearly on energy, while in the middle and at the end of the observation, the PD was constant with energy. 
Some of the previous studies of the WMNSs with \ixpe reported a rapid switch of the PA by up to $60\degr$, but the persistent rotation of the polarization plane was presented here for the first time. 

We suggested several ways to interpret the results of the data analysis. 
We studied the correlation of the dip in the light curve with the polarimetric properties and concluded that although the dip corresponds to the anomaly in the PA and PD behavior in the dip-core part of the observation, it is not correlated with the general trends of the polarimetric behavior. 
We defined a constant polarized component and subtracted it from all the observations. We concluded that the second, time-variable component then is expected to switch the sign of the PD (i.e., the PA rotates by 90\degr) in the middle of the observation. 
We also discussed the possible impact of the wind and the unusual state in which we found the source. 
The observed variations in the PA likely imply a misalignment between the NS spin and the orbital axis. 
Future polarimetric observations will help us to uncover the nature of this unique phenomenon. 

%========================================
%========================================
\begin{acknowledgements}
The Imaging X-ray Polarimetry Explorer (IXPE) is a joint US and Italian mission.  The US contribution is supported by the National Aeronautics and Space Administration (NASA) and led and managed by its Marshall Space Flight Center (MSFC), with industry partner Ball Aerospace (contract NNM15AA18C).  The Italian contribution is supported by the Italian Space Agency (Agenzia Spaziale Italiana, ASI) through contract ASI-OHBI-2022-13-I.0, agreements ASI-INAF-2022-19-HH.0 and ASI-INFN-2017.13-H0, and its Space Science Data Center (SSDC) with agreements ASI-INAF-2022-14-HH.0 and ASI-INFN 2021-43-HH.0, and by the Istituto Nazionale di Astrofisica (INAF) and the Istituto Nazionale di Fisica Nucleare (INFN) in Italy.  This research used data products provided by the IXPE Team (MSFC, SSDC, INAF, and INFN) and distributed with additional software tools by the High-Energy Astrophysics Science Archive Research Center (HEASARC), at NASA Goddard Space Flight Center (GSFC). 
We thank the \xmm Project Scientist for approving our DDT request to observe \source.
The research leading to these results has received funding from the European Union’s Horizon 2020 Programme under the AHEAD2020 project (grant agreement 871158). 
This research has been supported by the Academy of Finland grants 333112, 349144, and 355672 (AB, SVF, JP, VL, AV,  APN, SST) and the German Academic Exchange Service (DAAD) travel grant 57525212 (VD).
ADM, FLM, SF, FMu, ECo,  RF, and PSo are partially supported by MAECI with grant CN24GR08 “GRBAXP: Guangxi-Rome Bilateral Agreement for X-ray Polarimetry in Astrophysics”.
This research was also supported by the INAF grant 1.05.23.05.06: ``Spin and Geometry in accreting X-ray binaries: The first multifrequency spectro-polarimetric campaign''. 
IL was supported by the NASA Postdoctoral Program at the Marshall Space Flight Center, administered by Oak Ridge Associated Universities under contract with NASA.
\end{acknowledgements}

\bibliographystyle{yahapj}
\bibliography{ixpe_publ}

\end{document}